\documentclass[conference]{IEEEtran}
\IEEEoverridecommandlockouts
\usepackage{cite}
\usepackage{amsmath,amssymb,amsfonts}
\usepackage{algorithmic}
\usepackage{graphicx}
\usepackage{textcomp}
\usepackage[dvipsnames]{xcolor}
\usepackage{cite}
\usepackage{amsrefs}
\usepackage{siunitx}
\usepackage[font=footnotesize]{caption}
\usepackage{subcaption}
\usepackage{stackengine}
\usepackage{tikz}
\usepackage[siunitx]{circuitikz}
\usepackage[colorlinks=true,allcolors=blue]{hyperref}
\usepackage{soul}

\def\BibTeX{{\rm B\kern-.05em{\sc i\kern-.025em b}\kern-.08em
    T\kern-.1667em\lower.7ex\hbox{E}\kern-.125emX}}
\begin{document}

\title{Exploration of superconducting multi-mode cavity architectures for quantum computing\\

\thanks{This material is based upon work supported by the U.S. Department of Energy, Office of Science, National Quantum Information Science Research Centers, Superconducting Quantum Materials and Systems Center (SQMS) under contract number DE-AC02-07CH11359.}
}

\author{
\IEEEauthorblockN{Alessandro Reineri}
\IEEEauthorblockA{\textit{Fermi National Accelerator Laboratory} \\
Batavia, USA \\
areineri@fnal.gov}
\and
\IEEEauthorblockN{Silvia Zorzetti}
\IEEEauthorblockA{\textit{Fermi National Accelerator Laboratory} \\
Batavia, USA \\
zorzetti@fnal.gov}
\and
\IEEEauthorblockN{Tanay Roy}
\IEEEauthorblockA{\textit{Fermi National Accelerator Laboratory} \\
Batavia, USA \\
roytanay@fnal.gov}
\and
\IEEEauthorblockN{Xinyuan You}
\IEEEauthorblockA{\textit{Fermi National Accelerator Laboratory} \\
Batavia, USA \\
xinyuan@fnal.gov}
}

\maketitle

\begin{abstract}
Superconducting radio-frequency (SRF) cavities coupled to transmon circuits have proven to be a promising platform for building high-coherence quantum information processors. An essential aspect of this realization involves designing high quality factor three-dimensional superconducting cavities to extend the lifetime of quantum systems. To increase the computational capability of this architecture, we are exploring a multi-mode approach. This paper presents the design optimization process of a multi-cell SRF cavity to perform quantum computation based on an existing design developed in the scope of particle accelerator technology. 
We perform parametric electromagnetic simulations to evaluate and optimize the design. 
In particular, we focus on the analysis of the interaction between a nonlinear superconducting circuit known as the transmon and the cavity. This parametric design optimization is structured to serve as a blueprint for future studies on similar systems.
\end{abstract}


\section{Introduction}
\label{sec:Intro}
Superconducting radio-frequency (SRF) cavity and transmon coupled systems have the potential to be a primary architecture for the realization of high-coherence quantum information processors \cite{Reinhold}. Moreover, such systems show good scalability toward the metric of quantum volume \cites{Cross,Jurcevic}. The information is typically encoded in the lowermost $N$ states of a cavity mode, forming a more complex object called ``qudit''. The control of the cavity states is performed by the transmon \cite{Koch}, a superconducting nonlinear oscillator. One way of scaling up the amount of information encoded in these systems is to use higher order modes of the cavity, which usually show a lower coherence time than the fundamental mode, because of the difference in the modes' quality factor. A better approach is to use a multi-cell cavity containing several modes with nearly the same quality factor, where the mentioned modes are found within a bandwidth. Multi-cell cavities were originally developed in high-energy physics for particle acceleration purposes. Consisting of a repetition of the single-cell cavity geometry, the multi-cell cavity is an intrinsic multi-modal resonator with a number of high quality factor modes equal to the number of cells. Additionally, such a resonator geometry allows for a bigger Hilbert space to implement more complex quantum algorithms.
\par At Fermilab, the superconducting multi-cell architecture has been studied and manufactured since the year 2000 \cite{PhysRevSTAB.3.092001}. Among the most widespread multi-cell designs, the TESLA-shape cavity has shown astonishing $Q_{0}$ values for all its fundamental modes, up to $10^{10}$ \cite{alam2022quantum}, where $Q_{0}$ stands for the cavity's internal quality factor, i.e. the number of oscillations a cavity mode's electric or magnetic field undergoes before being dissipated. However, since the mentioned cavity geometry has been developed for particle acceleration purposes, the strength of interaction between transmon and different cavity modes varies by orders of magnitude. This ultimately poses a major challenge in controlling the transmon-cavity coupled system with the available driving protocols \cites{Heeres_2015,Eickbusch2022}. Therefore, the multi-cell cavity shape has to be modified to decrease the difference in transmon-mode interactions among all the cavity fundamental $\mathrm{TM}_{010}$ modes. Yet, given the novelty of the architecture, there is no established cavity design optimization process for multi-mode resonator shapes to be used in the field of quantum computation.
\par In this paper, we present an example of multi-mode cavity design along with the optimization workflow. The goal is to find the most suitable design for quantum computation purposes. The optimization, based on finite-element electromagnetic simulations implemented with the software CST Studio Suite\textsuperscript{\textregistered} and Ansys\textsuperscript{\textregistered} High-Frequency Structure Solver (HFSS\textsuperscript{\texttrademark}), aims to determine the cavity geometric parameters of interest and their influence on the cavity-transmon interaction strength for each fundamental eigenmode. At first, by carefully varying the identified parameters, the workflow allows us to obtain a multi-cell resonator shape that meets the desired requirement. Then, we evaluate the interaction between the fields and the transmon by inserting a representative sample of a transmon chip inside the cavity and performing a second series of electromagnetic simulations. 
\par For a better explanation of the design optimization workflow we also include in the paper a preliminary example of multi-cell cavity design developed using the shown process. Specifically, the presented design is found starting from the aforementioned, already-established TESLA-shape structure. We chose to limit our investigation to a relatively small 3-cell cavity to keep the number of modifiable geometric parameters low, though the process can be generalized to resonators with more complex geometry.
\par The cavity-transmon interaction parameters we find with this second set of simulations are compatible with some of the available control protocols \cites{Heeres_2015,Eickbusch2022}. Moreover, the mentioned parameters' ranges also cover the values recently obtained in an experimental study using the same transmon positioning method, though with a different cavity design \cite{Milul}.

\section{The TESLA-shape geometry}
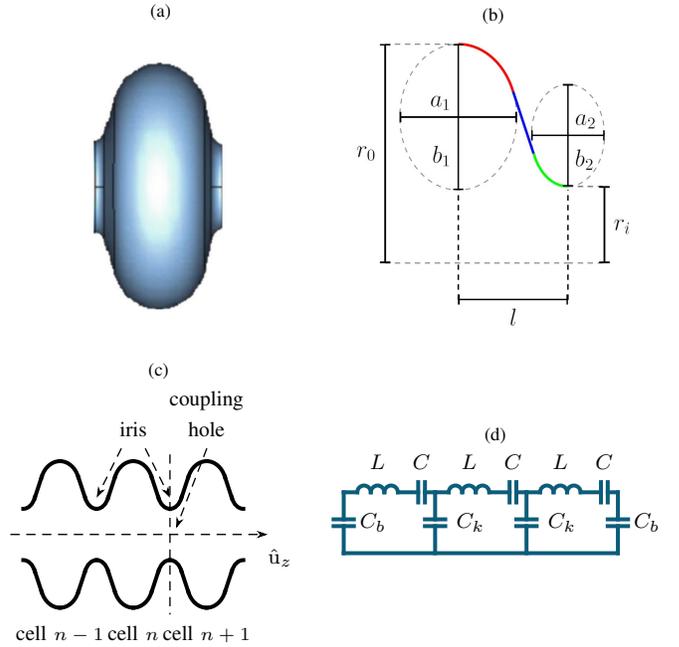
\begin{figure}[htbp]
    \centering
    \begin{subfigure}[c]{0.237 \textwidth}
        \centering
        \caption{}
        \includegraphics[scale = 0.18]{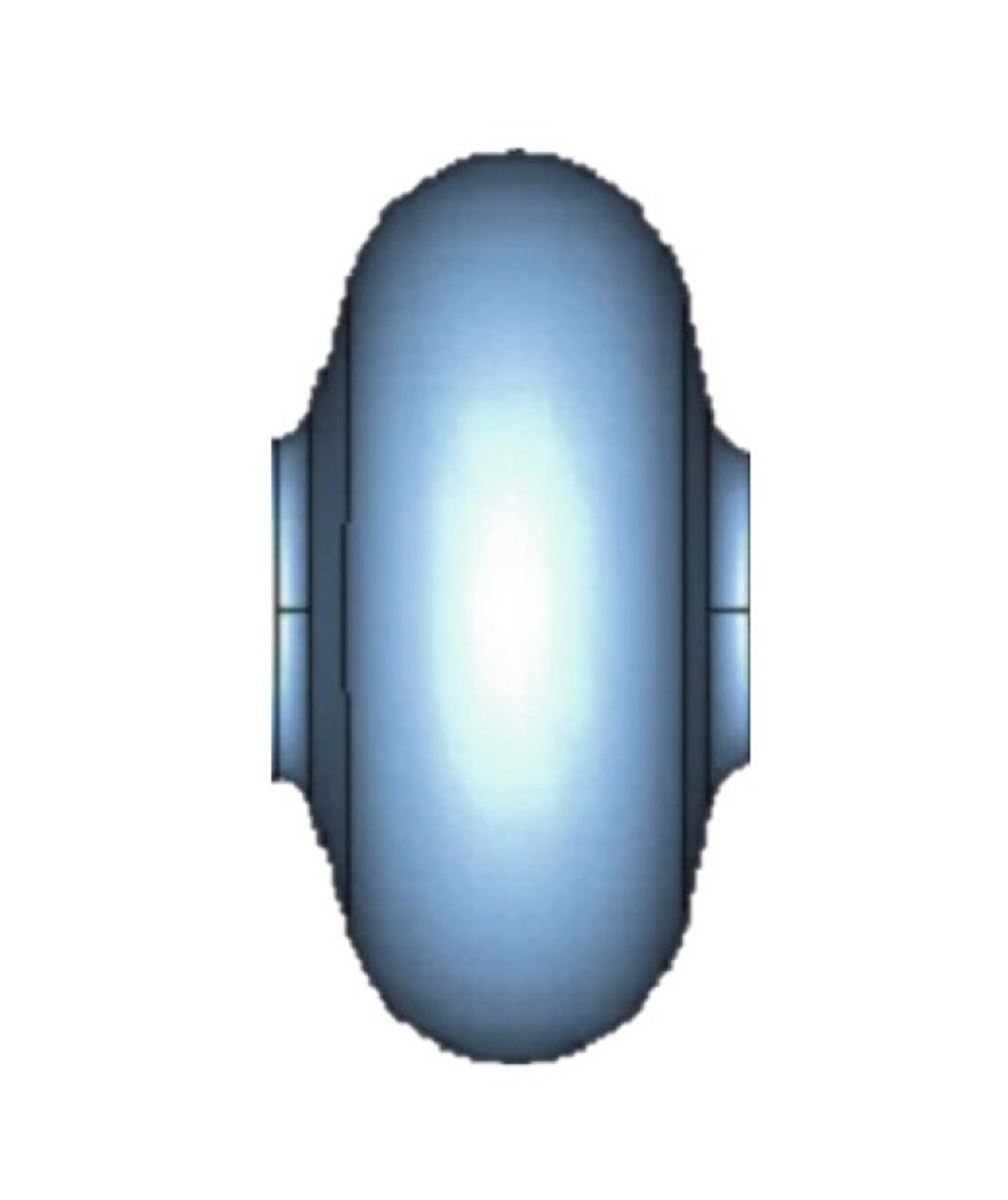}
        \label{fig:SingCellCAD}
    \end{subfigure}
    \begin{subfigure}[c]{0.237 \textwidth}
        \centering
        \caption{}
        \includegraphics[scale = 0.17]{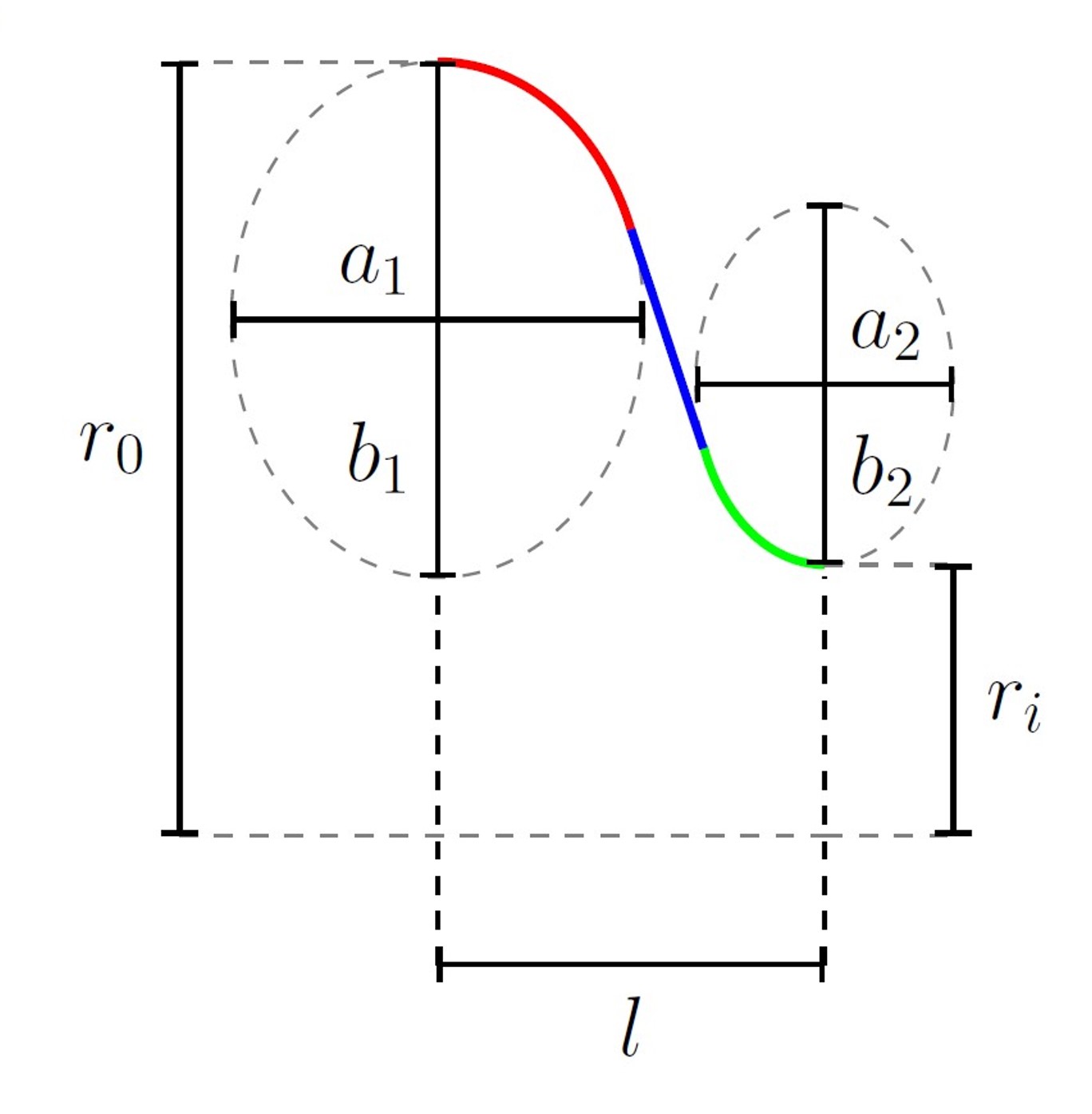}
        \label{fig:HalfCellGeomDiag}
    \end{subfigure}
    \begin{subfigure}[c]{0.24 \textwidth}
        \centering
        \caption{}
        \begin{tikzpicture}[scale = 0.325, line cap = rect]
            
             \foreach \x in {-1,0,1}
             {
               \draw [line width = 1.5] (3*\x,3) arc[start angle=90, end angle=20, x radius=0.8, y radius=1];
               \draw [line width = 1.5] (3*\x + 1.5,1.05) arc[start angle=-90, end angle=-160, x radius=0.5, y radius=0.7];
               \draw [line width = 1.5] (3*\x + 0.75,2.35) -- (3*\x + 1.03,1.5);
               \draw [line width = 1.5] (3*\x,3) arc[start angle=90, end angle=160, x radius=0.8, y radius=1];
               \draw [line width = 1.5] (3*\x - 1.5,1.05) arc[start angle=-90, end angle=-20, x radius=0.5, y radius=0.7];
               \draw [line width = 1.5] (3*\x - 0.75,2.35) -- (3*\x - 1.03,1.5);

               \draw [line width = 1.5] (3*\x,-3) arc[start angle=-90, end angle=-20, x radius=0.8, y radius=1];
               \draw [line width = 1.5] (3*\x + 1.5,-1.05) arc[start angle=90, end angle=160, x radius=0.5, y radius=0.7];
               \draw [line width = 1.5] (3*\x + 0.75,-2.35) -- (3*\x + 1.03,-1.5);
               \draw [line width = 1.5] (3*\x,-3) arc[start angle=-90, end angle=-160, x radius=0.8, y radius=1];
               \draw [line width = 1.5] (3*\x - 1.5,-1.05) arc[start angle=90, end angle=20, x radius=0.5, y radius=0.7];
               \draw [line width = 1.5] (3*\x - 0.75,-2.35) -- (3*\x - 1.03,-1.5);
             }
             
             \draw [line width = 0.5, dashed, -{Stealth[length=1.5mm,width=1mm]}] (-5,0) -- (5.5,0);
             \draw [line width = 0.5, -{Stealth[length=1.5mm,width=1mm]}, dashed] (-0.5,3.5) -- (-1.5,1.5);
             \draw [line width = 0.5, -{Stealth[length=1.5mm,width=1mm]}, dashed] (0.5,3.5) -- (1.5,1.5);
             \draw [line width = 0.5, dashed] (1.5,3.25) -- (1.5,-3.25); 
             \draw [line width = 0.5, dashed, -{Stealth[length=1.5mm,width=1mm]}]
               (3,3.5) -- (1.75,0.25);

             \node at (6,0)[below = 2pt]{\footnotesize $\mathrm{\hat{u}}_{z}$};
             \node at (0,3.5)[above = 1pt]{\footnotesize iris};
             \node at (3,3.5)[above = 1pt, align=center]{\footnotesize coupling\\ \footnotesize hole};
             \node at (-3,-3.25)[below = 1pt]{\footnotesize cell \scriptsize $n-1$};
             \node at (0,-3.25)[below = 1pt]{\footnotesize cell \scriptsize $n$};
             \node at (3,-3.25)[below = 1pt]{\footnotesize cell \scriptsize $n+1$};
            
        \end{tikzpicture}
        \label{fig:TESLAshapeCrossSec}
    \end{subfigure}
    \begin{subfigure}[c]{0.24 \textwidth}
        \centering
        \caption{}
        \begin{tikzpicture}[line width = 1.5, color = MidnightBlue!70!Black, scale = 0.18, line cap = rect]
            
            \foreach \r in {-1,0,1}
            {
               \draw (-0.5 + 6.75*\r,3) -- (0.5 + 6.75*\r,3);
            
               \foreach \t in {0,1,2}
               {
                  \draw (0.5 + \t + 6.75*\r,3) arc[start angle=180, end angle=0, radius=0.5];
               }
               \draw (3.5 + 6.75*\r,3) -- (5 + 6.75*\r,3);
               \draw (5 + 6.75*\r,3.75) -- (5 + 6.75*\r,2.25);
               \draw (5.5 + 6.75*\r, 3.75) -- (5.5 + 6.75*\r,2.25);
               \draw (5.5 + 6.75*\r, 3) -- (6.25 + 6.75*\r,3);
               \draw (-0.5 + 6.75*\r,3) -- (-0.5 + 6.75*\r,1);
               \draw (-0.5 + 6.75*\r, 0.5) -- (-0.5 + 6.75*\r, -1.5);
               \draw (-1.25 + 6.75*\r,1) -- (0.25 + 6.75*\r,1);
               \draw (-1.25 + 6.75*\r,0.5) -- (0.25 + 6.75*\r,0.5);
            }
               \draw (13, 3) -- (13, 1);
               \draw (13, 0.5) -- (13, -1.5);
               \draw (12.25,1) -- (13.75,1);
               \draw (12.25,0.5) -- (13.75,0.5);
               \draw (-7.25,-1.5) -- (13,-1.5);

               \foreach \p in {-1,0,1}
               {
                  \node at (2 + 6.75*\p,3.5)[above = 2pt, color = black]{\footnotesize $L$};
                  \node at (5.25 + 6.75*\p,3.5)[above = 2pt, color = black]{\footnotesize $C$};
               }
               \foreach \q in {0,1}
               {
                  \node at (-0.5 + 6.75*\q,0.75)[right = 4pt, color = black]{\footnotesize $C_{k}$};
               }
               \node at (-7.25,0.75)[right = 2pt, color = black]{\footnotesize $C_{b}$};
               \node at (13,0.75)[right = 2pt, color = black]{\footnotesize $C_{b}$};   
        \end{tikzpicture}
        \label{fig:TESLAShapeLCEM}
    \end{subfigure}
    \caption{TESLA-shaped SRF cavities. (a) A single-cell cavity. (b) The shape is determined by a set of geometric parameters defining the half-cell outline. (c) Individual cells are joined using irises to make a multi-cell cavity. (d) Electrical circuit model of the cavity, consisting of capacitively-coupled LC circuits.}
    \label{fig:TESLAShapeFigAndModels}
\end{figure}

The starting cavity geometry we considered is the multi-cell superconducting TESLA shape. It is based on the single-cell structure originally developed in the field of particle accelerator technology \cite{PhysRevSTAB.3.092001}. The single-cell outline is comprised of two ellipses' arcs joined together by their common tangent. The cell is then realized by rotating the contour around the beam axis and adding the resulting half-cell to another half-cell (Fig.\hspace{0.1cm}\ref{fig:SingCellCAD}\hspace{0.1cm}-\hspace{0.1cm}\ref{fig:HalfCellGeomDiag}). This particular cell shape allows for extremely high quality factor values, up to $10^{10}$ for the fundamental $\mathrm{TM_{010}}$ mode of a niobium single-cell cavity \cite{Romanenko}. The multi-cell cavity is then realized by combining together several single-cell structures through geometric interfaces called \textit{irises} (Fig.\hspace{0.1cm}\ref{fig:TESLAshapeCrossSec}).
\par From the electromagnetic point of view, the multi-cell cavity can be modeled as a series of LC circuits linked with coupling capacitances (Fig.\hspace{0.1cm}\ref{fig:TESLAShapeLCEM}). This way, the eigenvalues problem is analytically solved yielding the following expression for the $n$-th mode
\begin{equation}
    \label{eq:EigvalRelation}
    \left(\frac{\nu_{n}}{\nu_{0}}\right)^{2} = 1+2k_{cc}\left[1-\cos\left(\frac{n\pi}{N}\right)\right],
\end{equation}
where $\nu_{0}$ is the resonant frequency of the LC circuit representing a single cell, $N$ is the number of single LC elements and the constant $k_{cc}$, called \textit{cell-to-cell coupling}, is equal to the ratio of each LC circuit's capacitance over the coupling one $k_{cc}=\frac{C}{C_{k}}$. Equation\hspace{0.1cm}$\left(\ref{eq:EigvalRelation}\right)$ well describes the actual cavity's fundamental $\mathrm{TM_{010}}$ band: the single, high quality factor mode of the single-cell architecture splits into $N$ high quality factor modes, as many as the number of cells, which still show the same electric and magnetic field orientations of $\mathrm{TM_{010}}$ modes \cite{Padamsee}.
\par At the cavity level, the frequencies are determined by acting on the half-cell geometric parameters. In particular, the last mode's frequency, often referred to as $\pi$-mode, is related to the half-cell length by
\begin{equation}
    \label{eq:PiModeFreqVsLen}
    l=\frac{c}{2\nu_{\pi}},
\end{equation}
$c$ being the speed of light in vacuum. The $\pi$-mode frequency also determines the half-cell equatorial radius through the empirical relation
\begin{equation}
    \label{eq:R0VsNuPiEmp}
    r_{0} = \frac{c}{\nu_{\pi}}.
\end{equation}
On the other hand, the fundamental modes' bandwidth is related to the cell-to-cell coupling factor $k_{cc}$ via
\begin{equation}
    \label{eq:KccVsBandwidth}
    k_{cc} = \frac{\nu_{\pi}^{2}-\nu_{0}^{2}}{\nu_{0}^{2}}.
\end{equation}
This parameter, at the cavity level, gauges the amount of electromagnetic energy exchanged between each cell. Its magnitude depends on the geometric shape of the iris connecting two subsequent cells, i.e. on its radius $r_{i}$. For iris radius values of $0.4 \ r_{0}$, characterizing the TESLA-shape design, $k_{cc}$ is very small, around $0.02$. Consequently, the spectral width of the fundamental band of cavities operating in the GHz regime is limited to few tens of MHz. 

Moreover, the bandwidth remains constant as the number of cells is increased, resulting in more fundamental modes being inserted in the same frequency range, further reducing the modes' spacing. A very small frequency separation between the modes is not ideal: when resonantly driving one of the cavity modes, the same drive could also off-resonantly drive other modes, causing unwanted classical cross-talk \cite{Heeres_2015}. 
\par By inserting a single transmon inside a multi-cell TESLA-shape cavity it is possible to realize a multi-qudit processor. However, the coupling strengths between the transmon and the cavity modes show sizable differences in magnitude. This is due to the variation of the electric field magnitude among the modes at the ends of the cavity, where the transmon is usually placed. As the transmon interacts with the cavity through an electric dipole term, differences in the electric field component parallel to the transmon dipole moment are reflected and amplified into variations among the parameters characterizing each transmon-cavity mode interaction. The differences in the interaction strengths can even span orders of magnitude, resulting in the inability to drive all the eligible modes of the transmon-cavity coupled system with the available control protocols. 
\par Contrarily to the case of the bandwidth and modes' separation, there is no direct connection, in literature, between a target mode's electric field distribution within a portion of the cavity volume and a certain cavity geometric parameter. Nonetheless, through several simulation iterations, it has been noticed a slight dependence of the electric field variation toward the cell-to-cell coupling, as will be discussed in the next section.

\section{Cavity design optimization process}
\begin{figure}[htbp]
   \centering
   \begin{subfigure}[c]{0.24 \textwidth}
       \centering
       \caption{}
       \includegraphics[scale = 0.12]{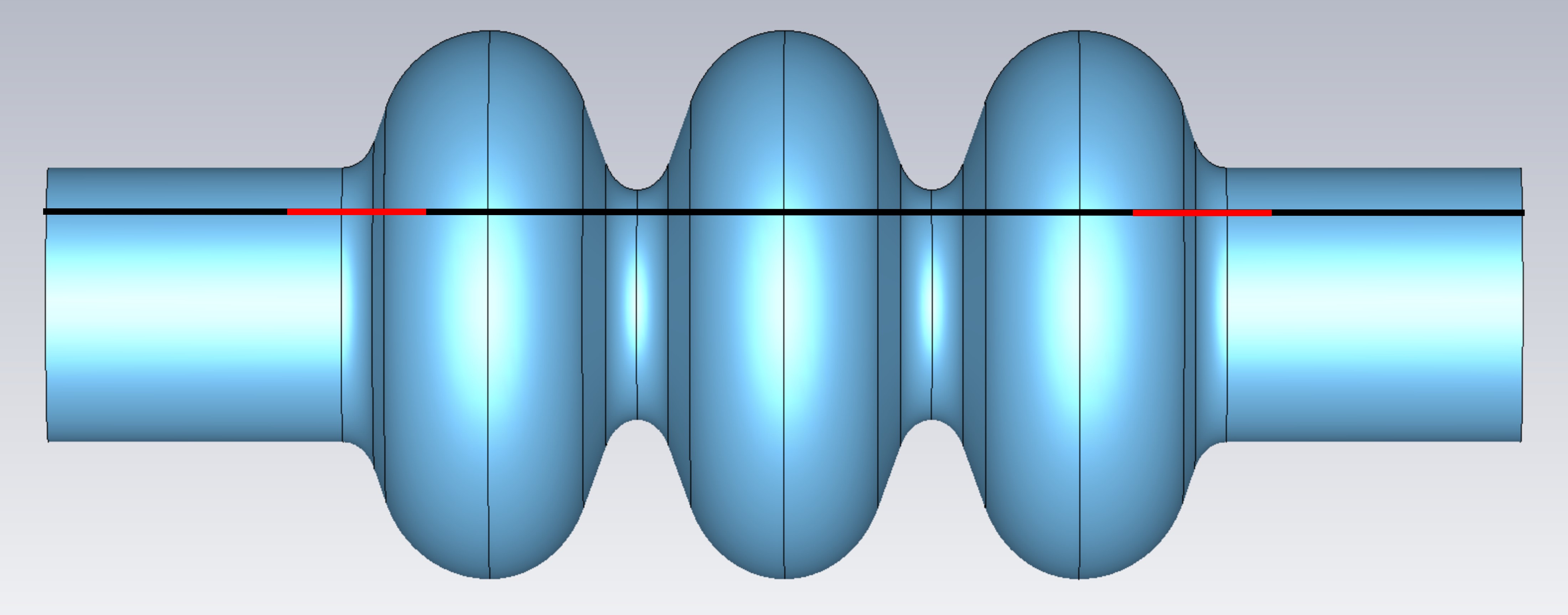}
       \label{fig:3CellTESLAShapeWithAxis}
   \end{subfigure}
   \begin{subfigure}[c]{0.24 \textwidth}
       \centering
       \caption{}
       \includegraphics[scale = 0.115]{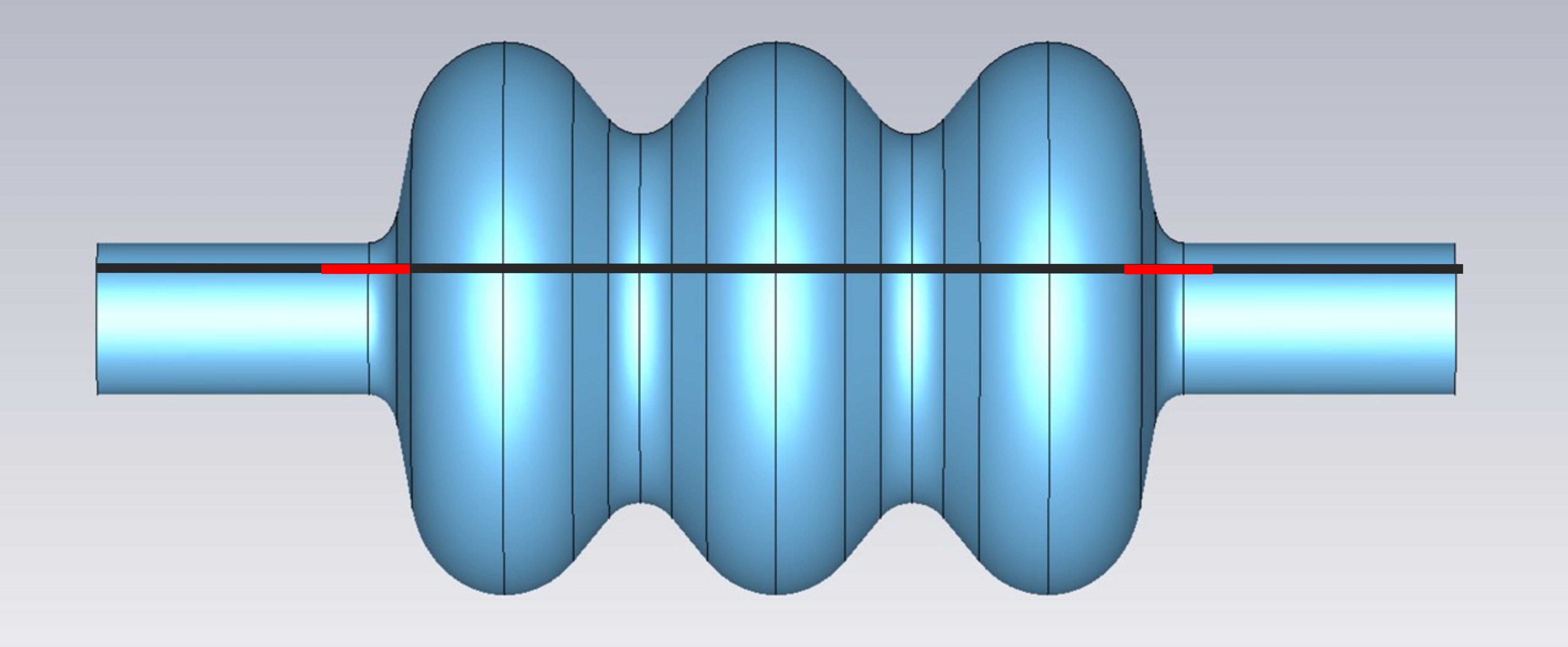}
       \label{fig:3CellTESLALikeWithAxis}
   \end{subfigure}
   \begin{subfigure}[c]{0.24 \textwidth}
       \centering
       \caption{}
       \includegraphics[scale = 0.32]{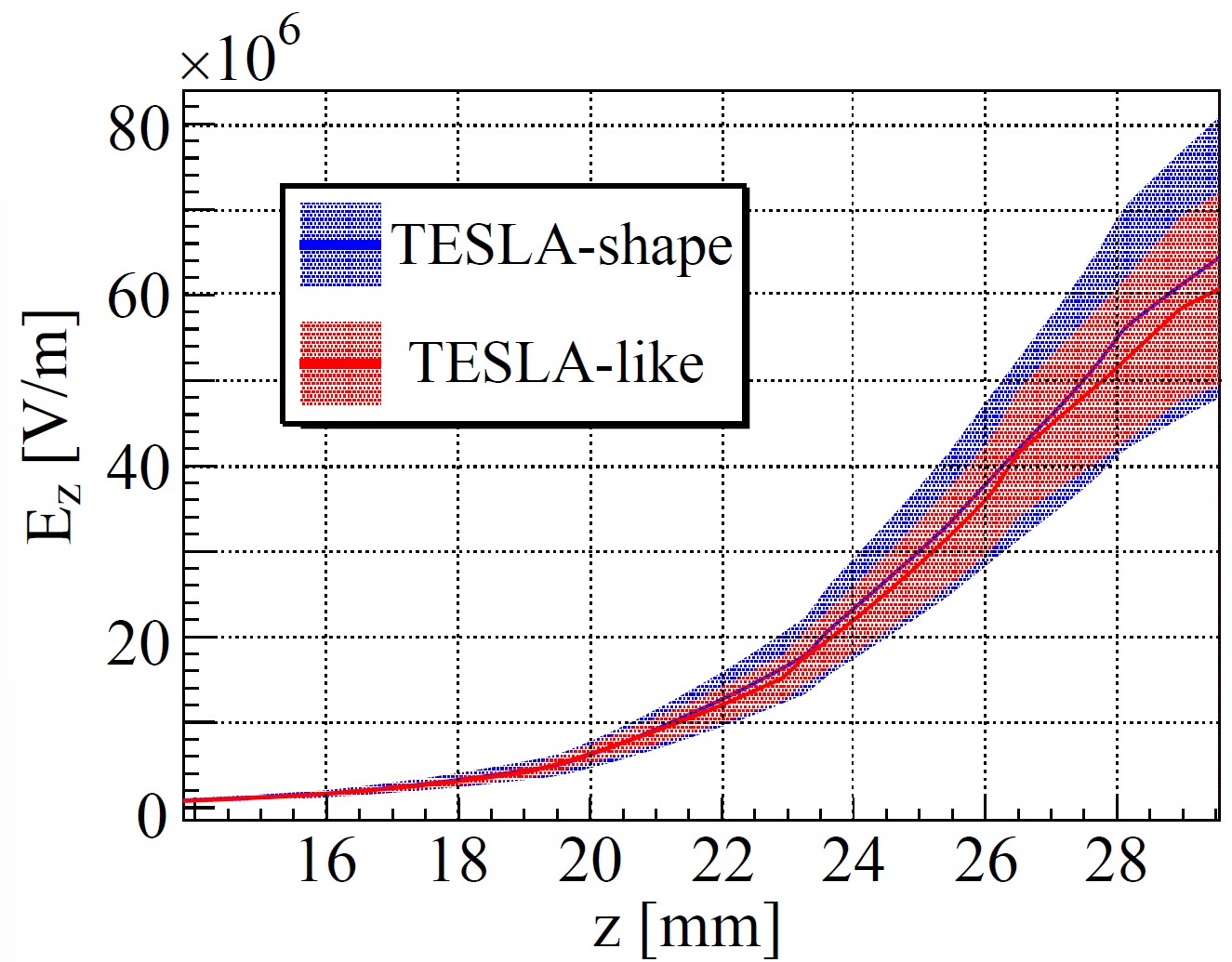}
       \label{fig:EzVsPosImprovLeftDet}
   \end{subfigure}
   \begin{subfigure}[c]{0.24 \textwidth}
       \centering
       \caption{}
       \includegraphics[scale = 0.32]{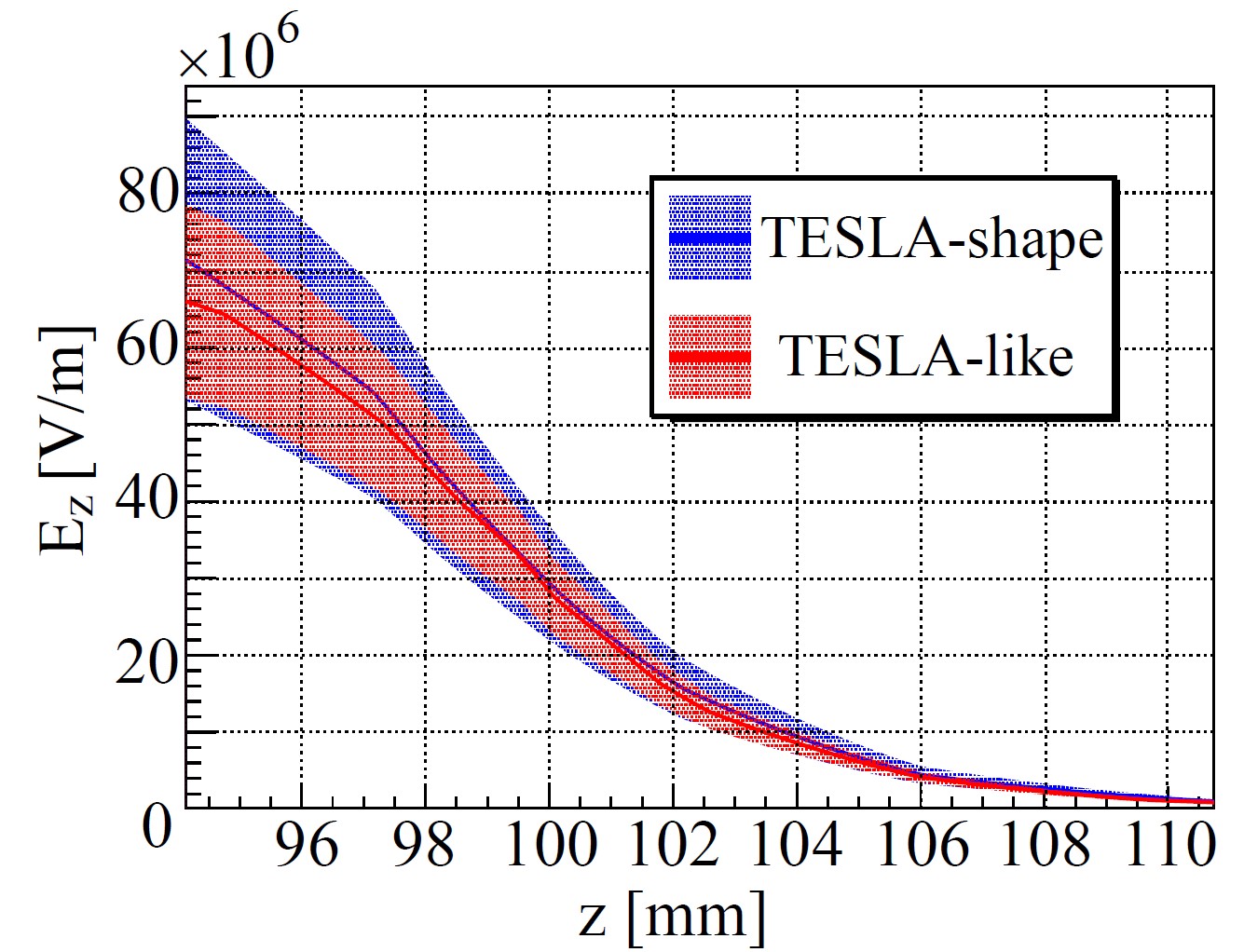}
       \label{fig:EzVsPosImprovRightDet}
   \end{subfigure}
   \caption{(a) TESLA-shape and (b) TESLA-like CAD models. The black lines with the red segments show the axis $l=\left(-4,0,z\right)$ along which the comparison of $E_{z}$ variation among the $\mathrm{TM}_{010}$ modes has been made. The plots (c) and (d) show that, although the $E_{z}$ mean value stays almost the same, there is a sizable reduction of the component variation in the optimized design case at both ends of the cavity, around 35\%.}
   \label{fig:3CellEzVarImprovVsDesign}
\end{figure}

In order to meet the aforementioned goals for the multi-cell cavity design improvement we started a design optimization process. The process is based on finite-element electromagnetic simulations involving the cavity alone and it is divided into two parts. In the first one, through the use of a CAD model of the cavity with all the half-cells parameterized independently, the simulations are set up with an eigenmode solver and prompted to compute the cavity's first $N$ eigenmodes for the considered geometry. Then, to study the effect of the iris radius modification on the bandwidth and modes' separation, the $r_{i}$ of all irises, all equal, are changed by the same amount and the eigenmode solver is rerun. In the second part of the optimization process, another series of eigenmode simulations are performed to evaluate the distribution of the electric field component parallel to the transmon dipole moment for all the $\mathrm{TM_{010}}$ modes. The analysis focuses on the neighborhood of the cavity entrances, along a line in which the transmon is typically inserted (Fig.\hspace{0.1cm}\ref{fig:3CellTESLAShapeWithAxis}\hspace{0.1cm}-\hspace{0.1cm}\ref{fig:3CellTESLALikeWithAxis}). Then, by modifying the two ellipses' semi-axis oriented in the $\mathrm{\hat{u}}_{z}$ direction by the same quantity for every half-cell, the electric field component is re-evaluated and compared to the former TESLA-shape design to assess eventual improvements in its variation among the modes. 
\par All these simulations are performed by considering the cavity walls made up of a perfect conductor, effectively not taking into account any electromagnetic loss. Additionally, the choice of modifying the geometric parameters of interest by the same quantity for all the half-cells is not a mandatory requirement. Indeed, it is possible to change each variable differently, all the half-cells being parameterized independently. However, having done so allows us to keep the cavity geometry simple and maintain the symmetry of the electric field distribution at both ends of the cavity, yet achieving the desired improvements. 
\par
\begin{figure}[htbp]
    \centering
    \begin{subfigure}[c]{0.24 \textwidth}
        \centering
        \caption{}
        \includegraphics[scale = 0.28]{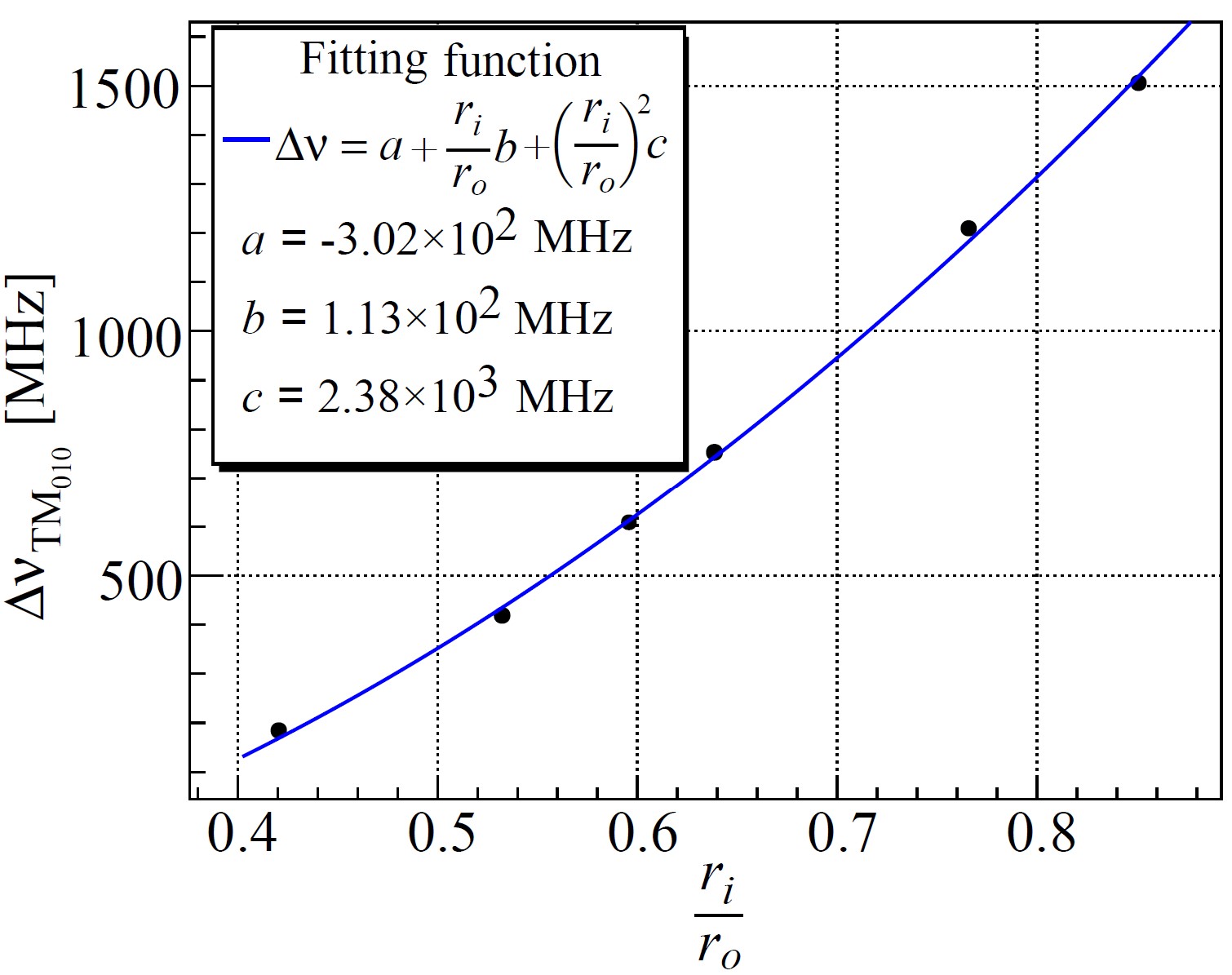}
        \label{fig:BandwidthVsRadiusRatio}
    \end{subfigure}
    \begin{subfigure}[c]{0.24 \textwidth}
        \centering
        \caption{}
        \includegraphics[scale = 0.28]{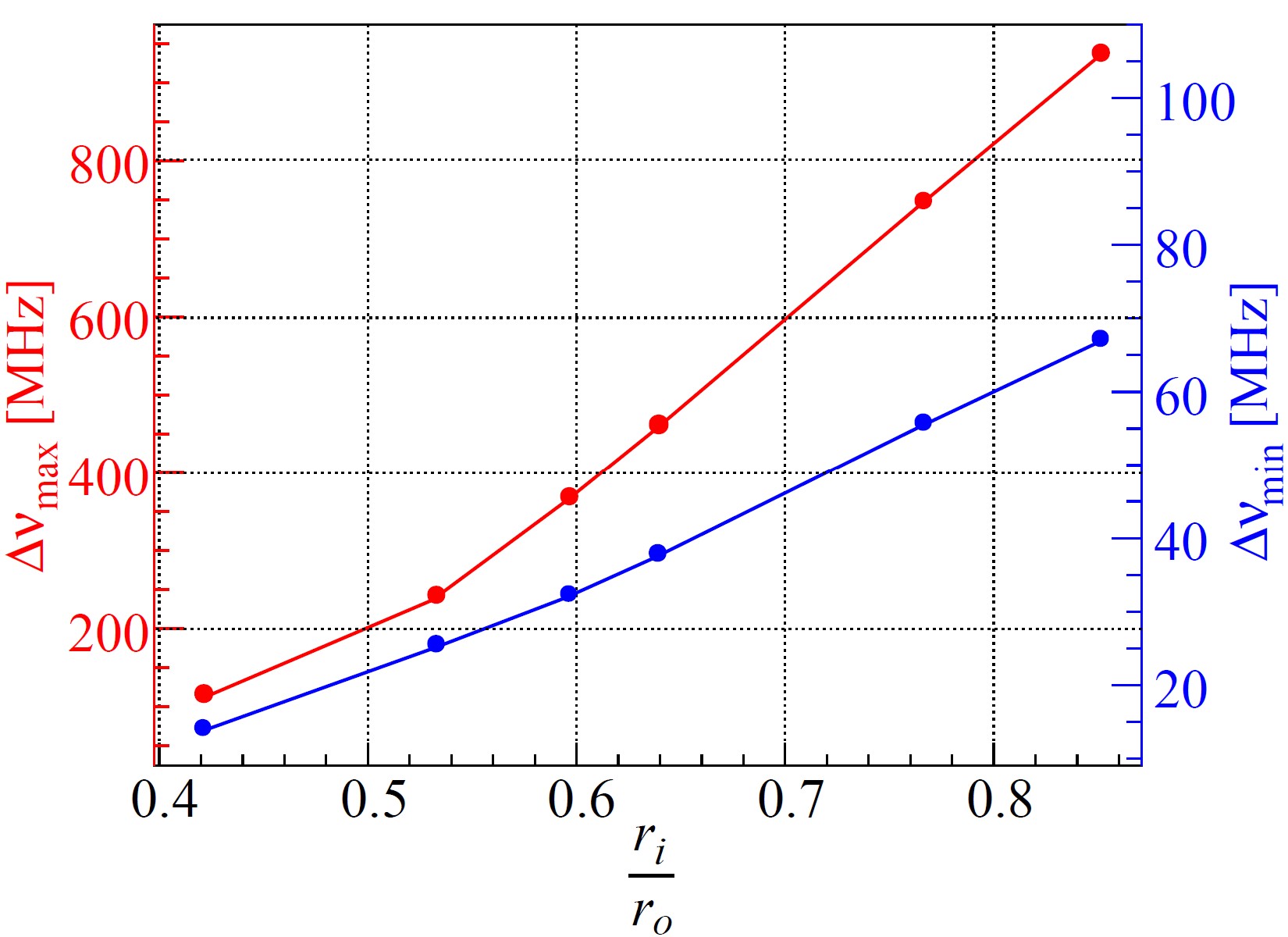}
        \label{fig:FreqDiffMinMaxVsRadiusRatio}
    \end{subfigure}
    \caption{$\mathrm{TM_{010}}$ bandwidth and inter-mode spacing as a function of the $\frac{r_{i}}{r_{0}}$ ratio. (a) The bandwidth shows a parabolic dependence. (b) Minimum and maximum modes' separation as a function of $\frac{r_{i}}{r_{0}}$.}
    \label{fig:BandWidthAndModesSepVsRadiusRatio}
\end{figure}

 We start the optimization process considering a three-cell TESLA-shape design sized to have a $\mathrm{TM_{010}}$ $\pi$-mode frequency of $\nu_{\pi}=6 \ \mathrm{GHz}$, i.e. with a half-cell length of $l=12.5 \ \mathrm{mm}$, an equatorial radius $r_{0}=23.55 \ \mathrm{mm}$, an iris radius $r_{i}=10 \ \mathrm{mm}$ and a radius ratio of $\frac{r_{i}}{r_{0}} = 0.42$. With this choice of parameters, the frequencies of the fundamental modes become 5.77 GHz, 5.88 GHz and 6.08 GHz, resulting in a 200 MHz wide $\mathrm{TM_{010}}$ passband with a maximum modes' separation of 110 MHz. By increasing the iris radius value relatively to $r_{0}$ and evaluating the fundamental band bandwidth, we notice that $\Delta\nu_{\mathrm{TM_{010}}}$ follows a parabolic relation towards $\frac{r_{i}}{r_{0}}$ (Fig.\hspace{0.1cm}\ref{fig:BandwidthVsRadiusRatio}). Consequently, with the number of modes remaining constant, the modes' separation increases as the iris is enlarged. In addition to that, for every iris radius value analyzed, all the fundamental $\mathrm{TM_{010}}$ modes keep their electromagnetic features, always showing the electric field oriented parallel to $\mathrm{\hat{u}}_{z}$. To not end up with a too-sharp and pointy cavity design that would affect its quality factor, we choose a radius ratio value of 0.63 which still provides a sizable improvement in the bandwidth, becoming 900 MHz large, and in the modes' separation, respectively of 270 MHz between the $\frac{\pi}{3}$ and the $\frac{2}{3}\pi$ modes and of 630 MHz between the $\frac{2}{3}\pi$ and the $\pi$-mode (Table\hspace{0.1cm}\ref{tab:TESLALike3CellBandData}).

\par For the second step of the optimization process, as mentioned before, we act on the parameters $r_{0}$ and $a_{2}$, slightly modifying them from the original values of the TESLA-shape design to see how the electric field's $z$-component distribution among the modes is affected. With the chosen radius ratio value from the first optimization step, prompting $r_{0} = 25.5 \ \mathrm{mm}$ for all the half-cells, $a_{2} = 2.7 \ \mathrm{mm}$ for the half-cells next to the beam stubs and $a_{2} = 3.8 \ \mathrm{mm}$ for the other half-cells, we notice a reduction of the 35\% in the $E_{z}$ variation among all the $\mathrm{TM_{010}}$ modes in the designated neighborhoods of the cavity entrances along the line $l=\left(-4,0,z\right)$ compared to the initial TESLA-shape design (Fig.\hspace{0.1cm}\ref{fig:3CellEzVarImprovVsDesign}). We refer to the optimized design with \textit{TESLA-like} for its resemblance to the original TESLA-shape one.

\par
\begin{figure}[htbp]
    \centering
    \includegraphics[scale = 0.17]{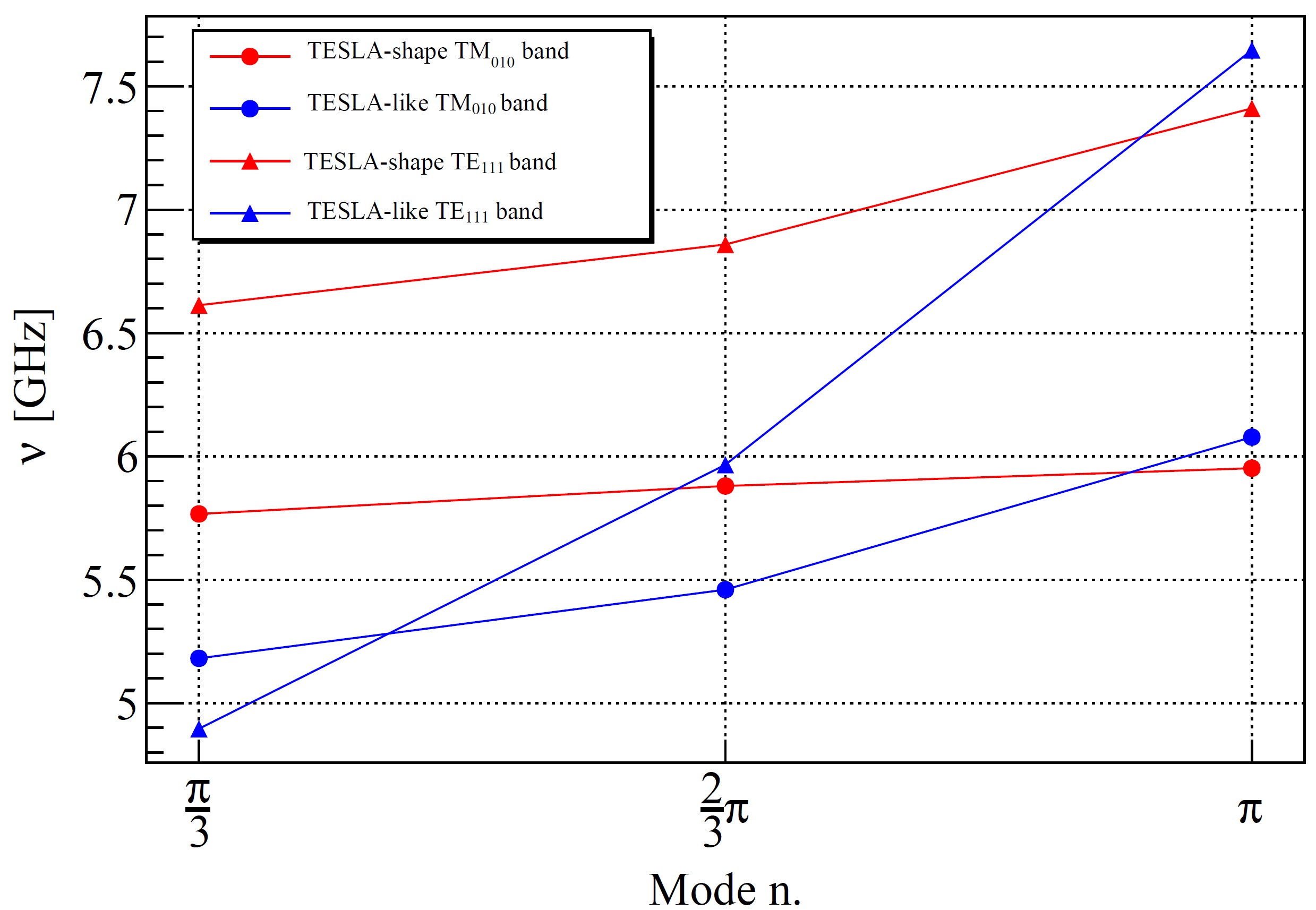}
    \caption{Band diagram comparison between TESLA-shape and TESLA-like cavity designs. The $\mathrm{TM_{010}}$ and $\mathrm{TE_{111}}$ bands, well separated in the first case, intersect one the other in the second case.}
    \label{fig:3CellCavitybandComp}
\end{figure}

 In modifying the cavity geometry, not only the $\mathrm{TM_{010}}$ band and modes are affected but also the higher order $\mathrm{TE_{111}}$ ones. In particular, the two bands, while being fairly apart in the case of the initial TESLA-shape design, intersect one another in the case of the TESLA-like design, resulting in the $\mathrm{TE_{111}}$ $\frac{2}{3}\pi$-mode to be found between the $\mathrm{TM_{010}}$ $\frac{2}{3}\pi$-mode and the $\pi$-mode (Fig.\hspace{0.1cm}\ref{fig:3CellCavitybandComp}). Despite that, the unwanted $\mathrm{TE_{111}}$ modes are not expected to interfere with the correct functioning of the cavity when the transmon is inserted. That is both due to a still fairly large separation between the $\mathrm{TE_{111}}$ $\frac{2}{3}\pi$ mode and the nearest $\mathrm{TM_{010}}$ mode, around 100 MHz (Table\hspace{0.1cm}\ref{tab:TESLALike3CellBandData}), and to the fact that, being transverse-electric modes, their electric field is oriented almost-perpendicularly to the transmon dipole. There might be a small $E_{z}$ component that couples with the transmon, for the cavity geometry makes the fields bend slightly in the vicinity of the cavity walls, though it is not expected to have a sizable effect on the system.

\begin{table}[htbp]
    \centering
    \caption{Band data comparison between TESLA-like and TESLA-shape designs.}
    \label{tab:TESLALike3CellBandData}
    \begin{tabular}{|c|c|c|}
         \hline
         \addstackgap[5pt]{\textbf{Mode}} & $\boldsymbol{\nu}$ \textbf{(TESLA-like) [GHz]} & $\boldsymbol{\nu}$ \textbf{(TESLA-shape) [GHz]}\\
         \hline
         \addstackgap[5pt]{$\mathrm{TM_{010}}$ $\frac{\pi}{3}$} & 5.1813 &  5.7666 \\
         \hline
         \addstackgap[5pt]{$\mathrm{TM_{010}}$ $\frac{2}{3}\pi$} & 5.4597 &  5.8780 \\
         \hline
         \addstackgap[5pt]{$\mathrm{TM_{010}}$ $\pi$} & 6.0780 &  5.9519 \\ 
         \hline
         \addstackgap[5pt]{$\mathrm{TE_{111}}$ $\frac{\pi}{3}$} & 4.8957 &  6.6127 \\
         \hline
         \addstackgap[5pt]{$\mathrm{TE_{111}}$ $\frac{2}{3}\pi$} & 5.9663 &  6.8589 \\ 
         \hline
         \addstackgap[5pt]{$\mathrm{TE_{111}}$ $\pi$} & 7.6470 &  7.4106 \\ 
         \hline
    \end{tabular}
\end{table}

\par
\begin{figure}[htbp]
    \centering
    \begin{subfigure}[c]{0.15 \textwidth}
        \centering
        \caption{}
        \includegraphics[scale = 0.08]{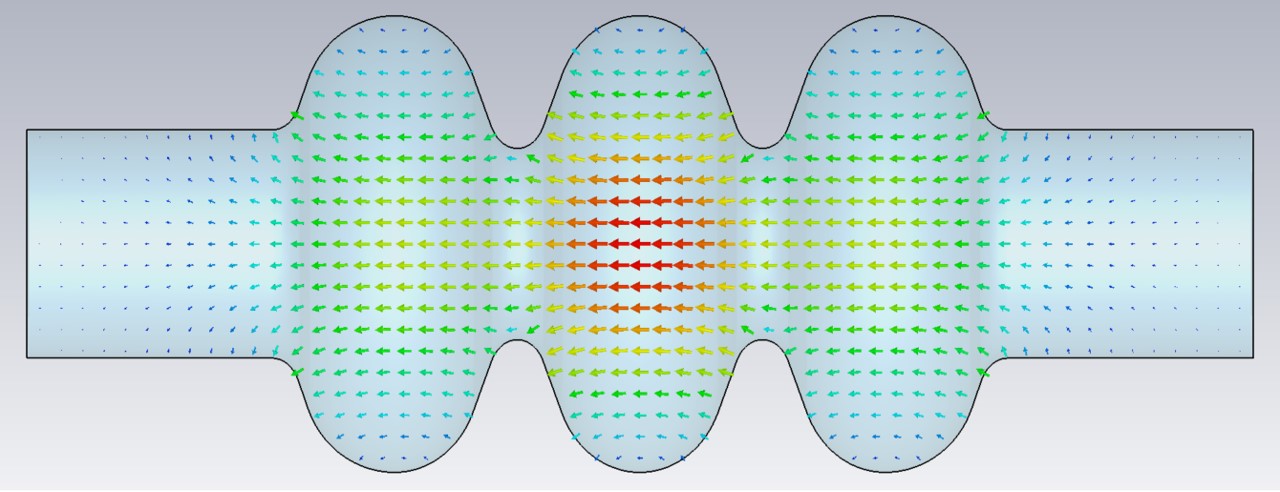}
        \label{fig:3cellTESLAshapeTMpi3}
    \end{subfigure}
    \begin{subfigure}[c]{0.15 \textwidth}
        \centering
        \caption{}
        \includegraphics[scale = 0.078]{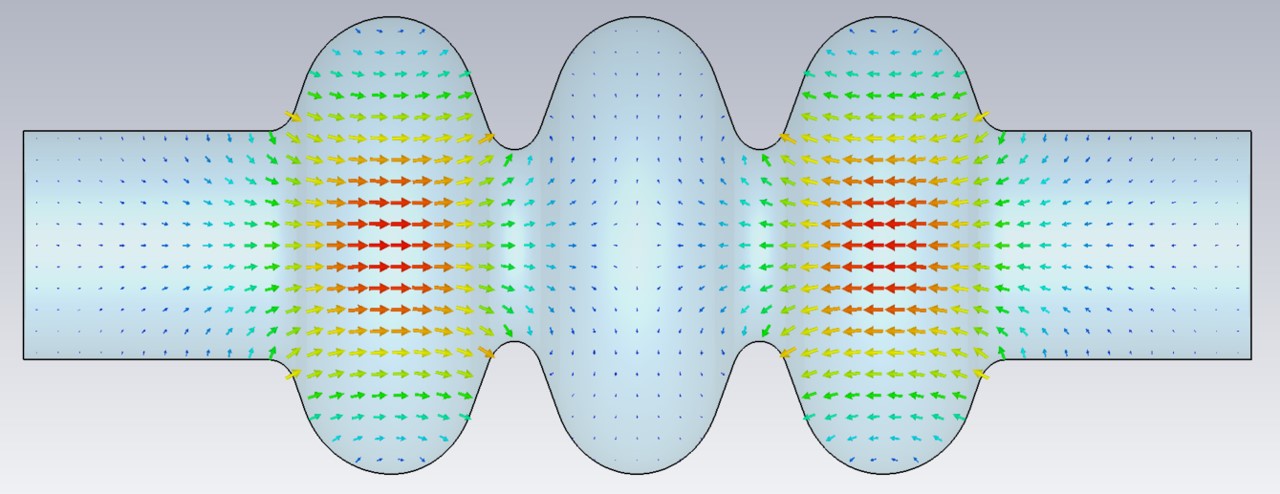}
        \label{fig:3cellTESLAshapeTM2pi3}
    \end{subfigure}
    \begin{subfigure}[c]{0.15 \textwidth}
        \centering
        \caption{}
        \includegraphics[scale = 0.08]{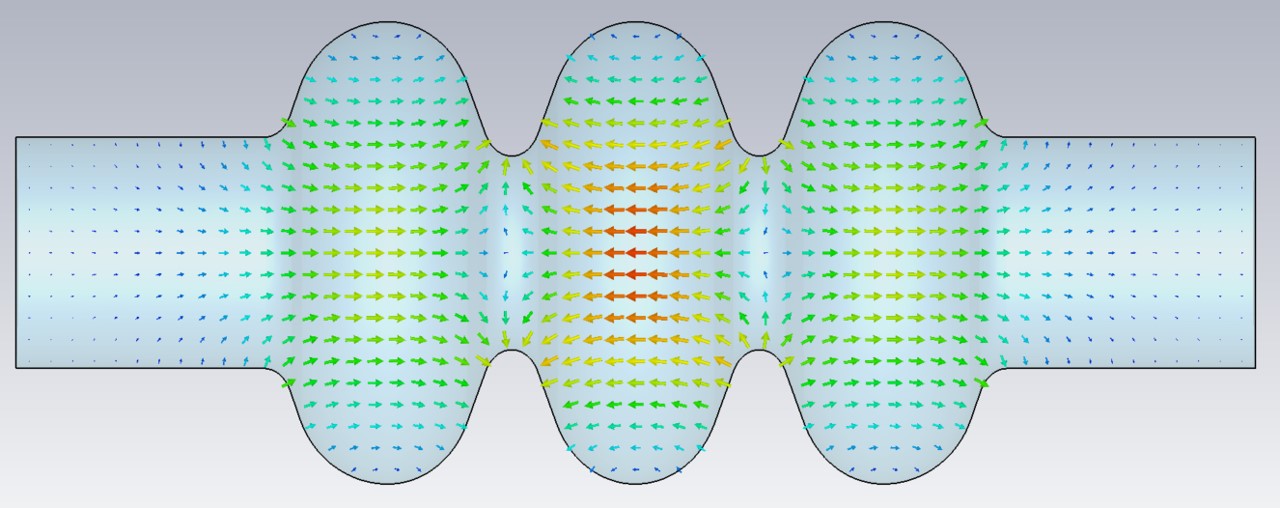}
        \label{fig:3cellTESLAshapeTMpi}
    \end{subfigure}
    \begin{subfigure}[c]{0.15 \textwidth}
        \centering
        \caption{}
        \includegraphics[scale = 0.08]{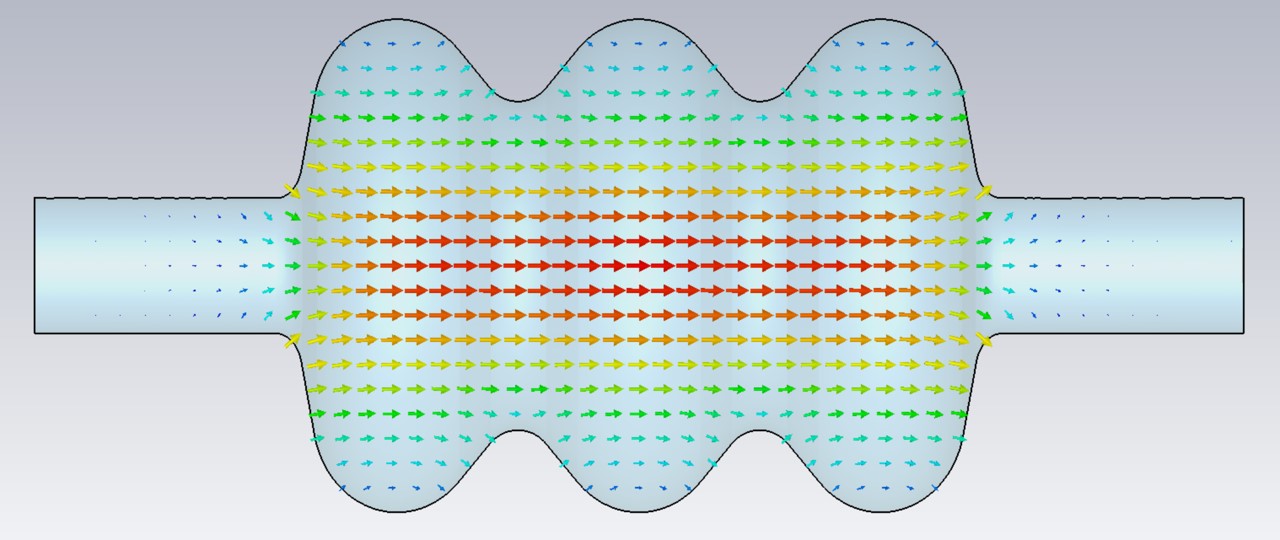}
        \label{fig:3cellTESLAlikeTMpi3}
    \end{subfigure}
    \begin{subfigure}[c]{0.15 \textwidth}
        \centering
        \caption{}
        \includegraphics[scale = 0.078]{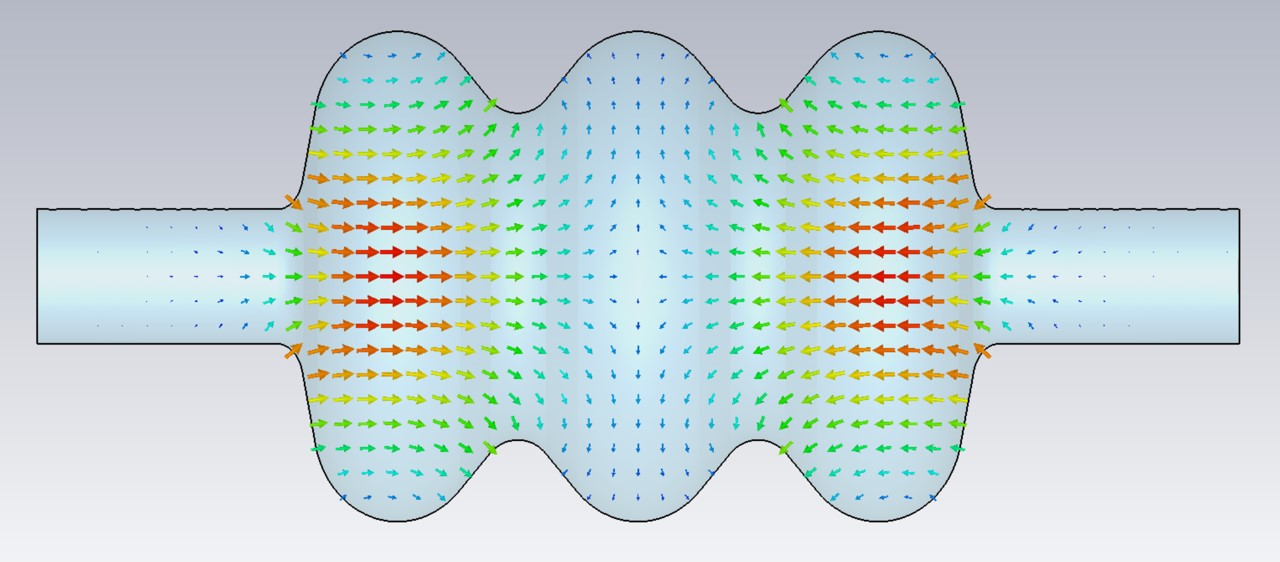}
        \label{fig:3cellTESLAlikeTM2pi3}
    \end{subfigure}
    \begin{subfigure}[c]{0.15 \textwidth}
        \centering
        \caption{}
        \includegraphics[scale = 0.08]{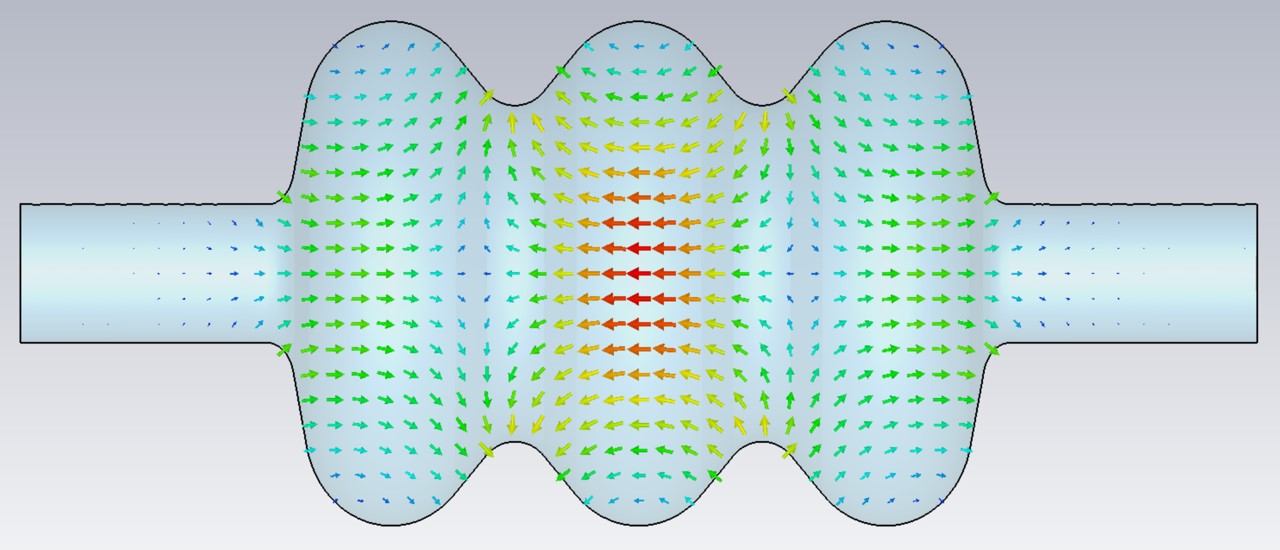}
        \label{fig:3cellTESLAlikeTMpi}
    \end{subfigure}
    \caption{Electric field distribution for the $\mathrm{TM_{010}}$ modes along x=0 plane. (a), (b), (c) show  the three fundamental modes for the TESLA-shape cavity and (d), (e), (f) show the same modes for the TESLA-like design. All the $\mathrm{TM_{010}}$ modes maintain their electromagnetic properties, i.e. their electric and magnetic field orientations, and the number of antinodes remain the same.}
    \label{fig:TESLALikeCADAndModesEPlot}
\end{figure}

 The found design meets the initial requirements of larger bandwidth and larger modes' separation. Moreover, we also obtain some improvement in the $E_{z}$ component variation among the modes. All the achievements did not modify the nature of the $\mathrm{TM_{010}}$ modes which still show their electric field original orientation. Figure\hspace{0.1cm}\ref{fig:TESLALikeCADAndModesEPlot} shows that the electric field orientation is the same along the $x=0$ cutting plane and that each mode keeps the number of field intensity maxima unchanged between the two designs.

\section{Transmon-cavity interaction evaluation}
\label{sec:CouplSysAnalysis}
The last step of the multi-cell cavity design optimization process involves assessing the interaction between the optimized TESLA-like design and the transmon. In particular, we want to evaluate all the parameters defining the Hamiltonian in the dispersive regime which, for an intrinsic multi-modal electromagnetic environment, reads \cite{Blais}
\begin{equation}
  \begin{split}
    \label{eq:MultiModeDispHam}
    \hat{H}_{\mathrm{disp}} \simeq & \sum_{m}\hbar\left(\omega_{m} + \gamma_{m}\right)\hat{a}_{m}^{\dag}\hat{a}_{m} \\ & + \frac{1}{2}\sum_{m}\hbar K_{m}\left(\hat{a}_{m}^{\dag}\right)^{2}\hat{a}_{m}^{2} \\ & + \sum_{m>n}\hbar\chi_{m,n}\hat{a}_{m}^{\dag}\hat{a}_{m}\hat{a}_{n}^{\dag}\hat{a}_{n},
  \end{split}
\end{equation}
where $\left\{\hat{a}_{m},\hat{a}_{m}^{\dag}\right\}_{m\in\mathbb{N}}$ are the $m$-th mode's annihilation and creation operators (including the transmon ones) and $\left\{\omega_{m}\right\}_{m\in\mathbb{N}}$ are the uncoupled system's eigenfrequencies (including the transmon one). The knowledge of the dispersive regime parameters $\left\{\gamma_{m}\right\}_{m\in\mathbb{N}}$, $\left\{K_{m}\right\}_{m\in\mathbb{N}}$ and $\left\{\chi_{m,n}\right\}_{m,n\in\mathbb{N},m\neq n}$, named respectively \textit{linear corrections}, \textit{self-Kerrs} and \textit{cross-Kerr} allows for a complete assessment of the transmon-cavity interaction in the dispersive regime working region \cite{Koch}. The parameters are not all independent one another; instead, the cross-Kerr interaction can be used to express the other quantities with the following
\begin{align}
    \gamma_{m} &= \frac{1}{2}\sum_{n}\chi_{m,n},  \label{eq:ParamRelationLinShifts} \\
    K_{m} &= \frac{\chi_{m,m}}{2}. \label{eq:ParamRelationSelfKerr}
\end{align}
In addition to that, the cross-Kerr interactions evaluation is fundamental to implementing quantum computation algorithms on a transmon-cavity coupled system \cites{Heeres_2015,Eickbusch2022}. A last, yet important parameter that does not pertain directly to the dispersive regime approximation is called \textit{Rabi coupling}. It gives a measure of the dipole interaction strength between each cavity $\mathrm{TM_{010}}$ mode and the transmon mode and can be calculated from the parameters in the equation (\ref{eq:MultiModeDispHam}) as \cite{Blais}
\begin{equation}
    \label{eq:GVsdispPars}
    g_{m} = \sqrt{\frac{-\chi_{0,m}\Delta_{m}\left(\Delta_{m}-\frac{E_{C}}{\hbar}\right)}{\frac{E_{C}}{\hbar}}}.
\end{equation}
where $\Delta_{m}$ is the frequency difference between the transmon and the $m$-th cavity mode, $\chi_{0,m}$ is the cross-Kerr interaction strength between the transmon and the $m$-th mode and $E_{C}$ is the transmon capacitive energy, related to the transmon self-Kerr interaction which, in literature, is called \textit{anharmonicity} \cite{Koch}.
\par
\begin{figure}[htbp]
    \centering
    \includegraphics[scale = 0.55]{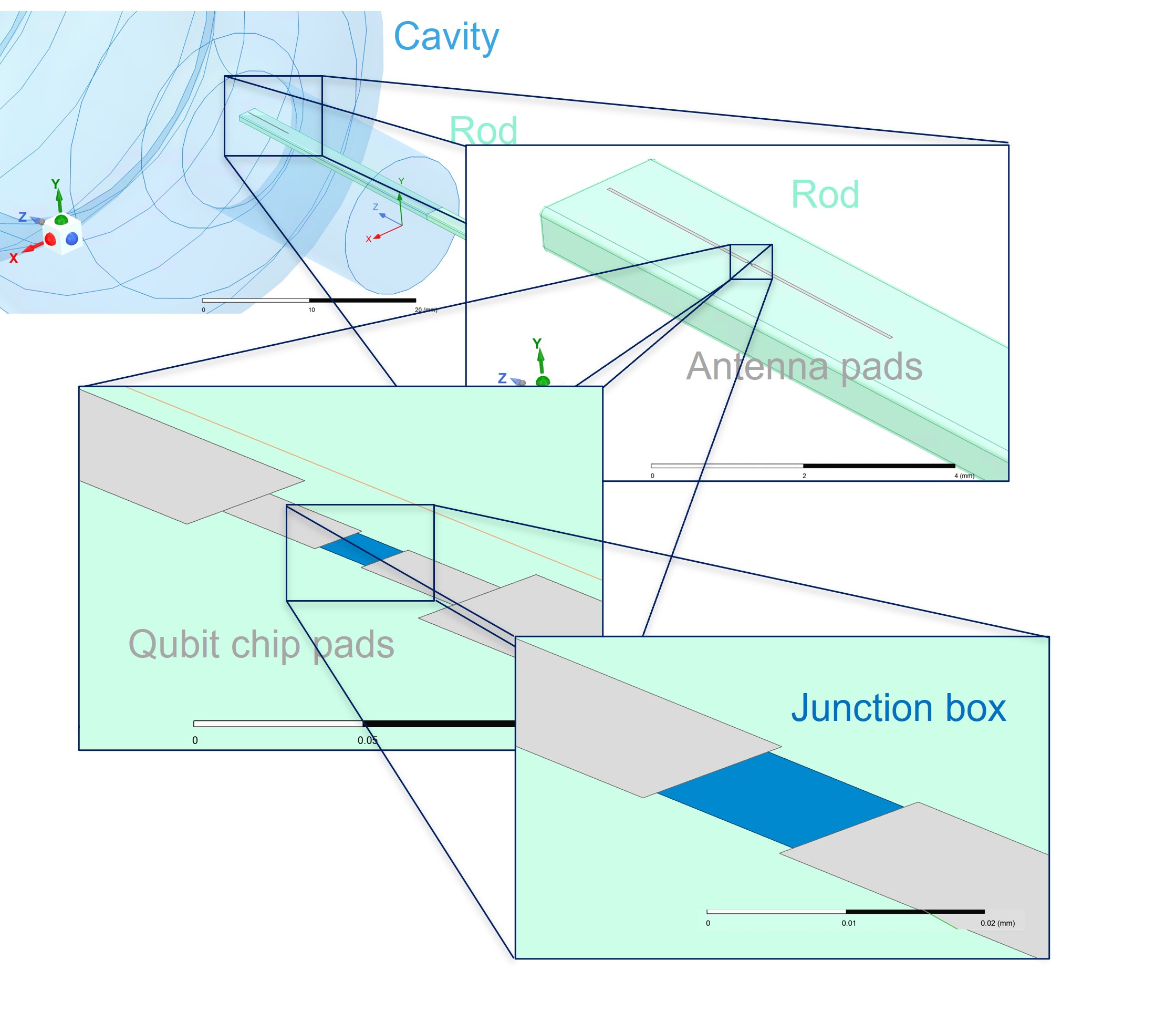}
    \caption{CAD model of the transmon inserted at one side of the cavity for EPR simulations.}
    \label{fig:3CellTESLAlikeAndTransmonEnsembleCAD}
\end{figure}

 To obtain all the mentioned parameters we use the energy-participation ratio analysis method \cite{Minev}. This protocol is based on finite-element eigenmode simulations and requires inserting a transmon element inside the cavity volume. It evaluates the fraction of electromagnetic energy stored in the transmon for each system's eigenmode from the $\mathbf{E}$ ad $\mathbf{H}$ distributions in the volume and, through that, returns all the parameters characterizing equation (\ref{eq:MultiModeDispHam}). The simulations are built by inserting a transmon at one end of the cavity on a loss-free silicon rod. The transmon geometry is similar to the ones already in literature and consists of a pair of antenna pads, a pair of qubit pads and a rectangle for the actual transmon junction, all modeled as perfect conductors (Fig.\hspace{0.1cm}\ref{fig:3CellTESLAlikeAndTransmonEnsembleCAD}). An additional boundary condition is set up on the junction rectangle by defining an integration line across it and assigning it a lumped inductance value that corresponds to the transmon linear Josephson inductance $L_{J_{0}}$. The chosen value $L_{J_{0}} = 11.57 \ \mathrm{nH}$ for this series of simulations corresponds to a transmon frequency of $\nu_{0} = 4.5 \ \mathrm{GHz}$.

\par Afterwards, we initialize the eigenmode solver with the number of desired eigenmodes we want the electric and magnetic fields to be calculated. At this stage, it is of the utmost importance to seed a mesh refinement both on the entire CAD model and locally around the transmon junction, for its small physical dimensions can make the solver not detect its impedance and result in failing the simulations. Finally, we set up a parametric sweep over the relative position between the cavity end and the transmon to see the behavior of the parameters around the neighborhood on which $E_{z}$ variation reduction among the modes is observed from the TESLA-shape to the TESLA-like design.
\par
\begin{figure}[htbp]
    \centering
    \includegraphics[scale = 0.32]{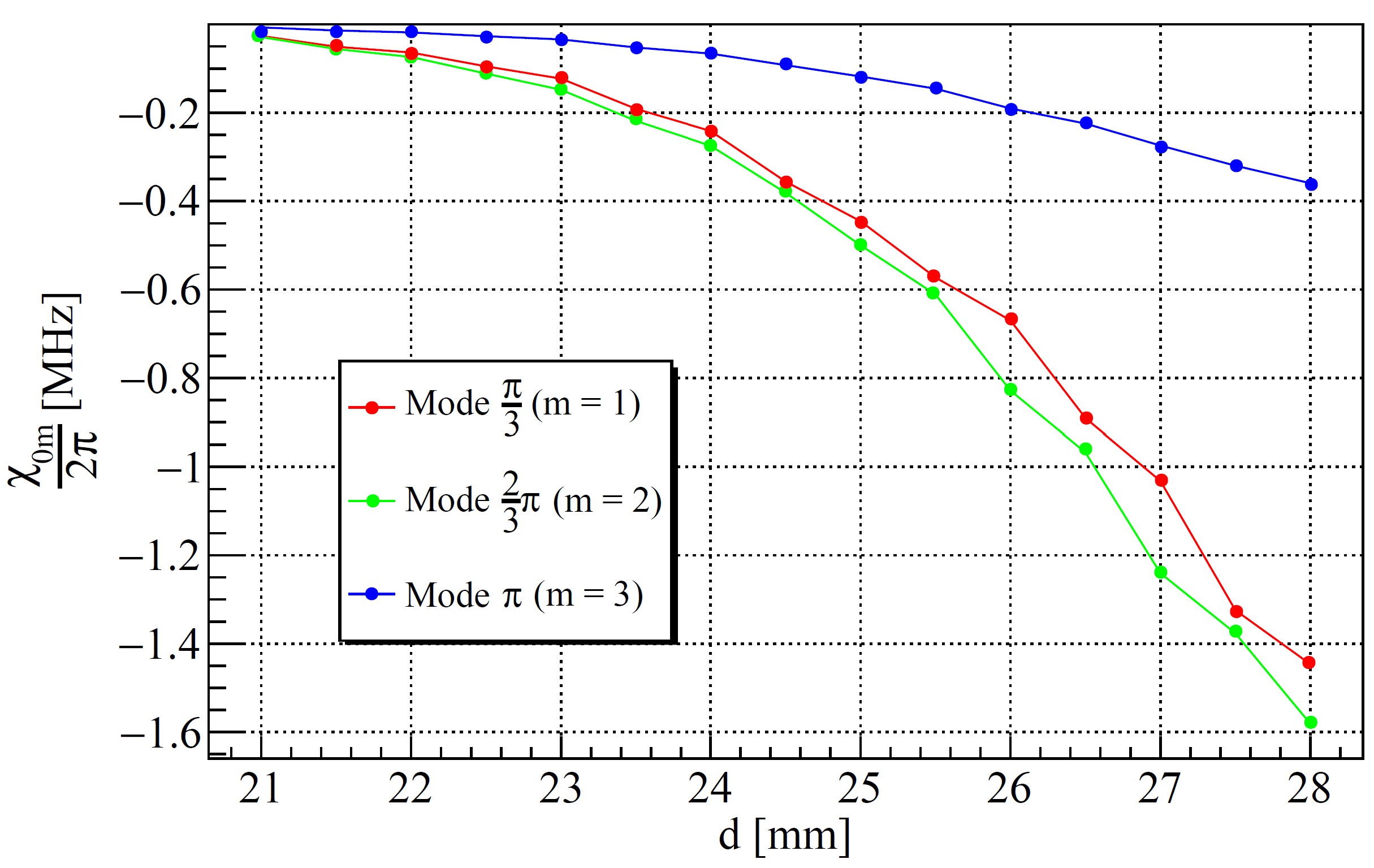}
    \caption{Cross-Kerr interaction between transmon and cavity modes as a function of transmon position for all cavity $\mathrm{TM_{010}}$ modes. The transmon position is given considering the system of coordinate origin at the beginning of the cavity beam stub so that larger values of $d$ imply the transmon being deeper inside the cavity.}
    \label{fig:3CellcrossKvsPosAllModes}
\end{figure}

 With the parameter values obtained from the cavity-transmon relative position sweep we plot Fig.\hspace{0.1cm}\ref{fig:3CellcrossKvsPosAllModes} showing each mode's parameter behavior toward the transmon position. 
Starting from the transmon-modes cross-Kerr interactions, we notice that the parameters $\chi_{01}$ and $\chi_{02}$ corresponding to the $\mathrm{TM_{010}}$ $\frac{\pi}{3}$-mode and $\frac{2}{3}\pi$-mode, are very close with each other for the considered transmon positions. 
The cross-Kerr $\chi_{03}$, instead, is much lower than the other two, about one order of magnitude less for each transmon position.  
As expected, all the cross-Kerr magnitudes decrease in absolute value as the transmon is moved out of the cavity, due to a smaller electric field. Moreover, all the modes' cross-Kerr remain negative within the entire position sweep. 
The performed sweep also shows that the cavity-transmon interaction can be tuned by repositioning the nonlinear circuit element with respect to the cavity entrance. 
This way, the coupled system can be rather flexible toward the applicable driving protocol by simply moving the transmon chip and retaining the multi-cell cavity design: the cross-Kerr interactions go from the order of MHz to tens of kHz, modifying the transmon-cavity interaction from strong dispersive regime to weak dispersive regime. This allows the system to be driven by either control protocols developed for strongly coupled \cite{Heeres_2015} or weakly coupled systems \cite{Eickbusch2022}.

\par
\begin{figure}[htbp]
    \centering
    \begin{subfigure}[c]{0.24 \textwidth}
       \centering
       \caption{}
       \includegraphics[scale = 0.325]{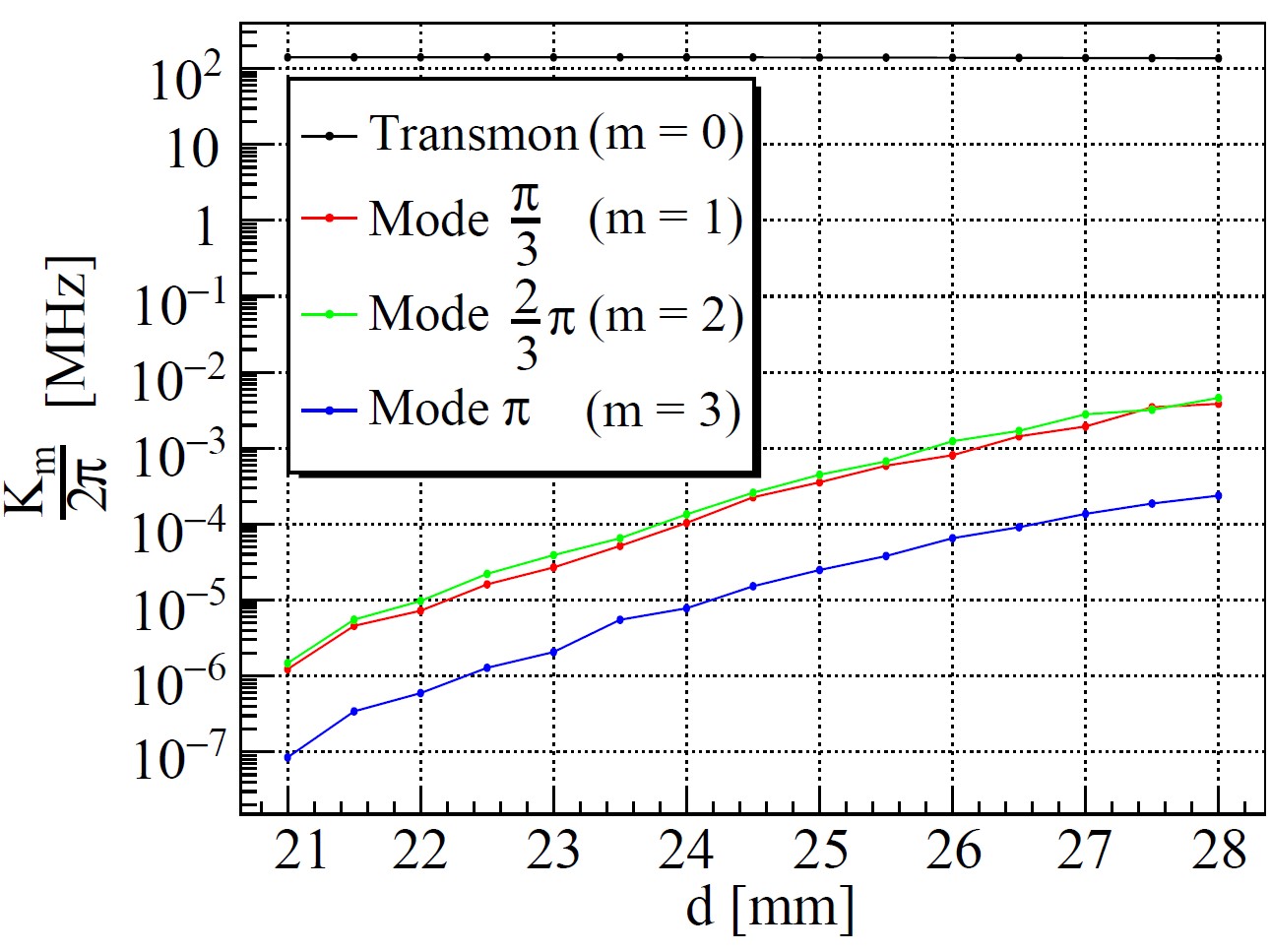}
       \label{fig:3CellselfKvsPosAllModes}
    \end{subfigure}
    \begin{subfigure}[c]{0.24 \textwidth}
       \centering
       \caption{}
       \includegraphics[scale = 0.265]{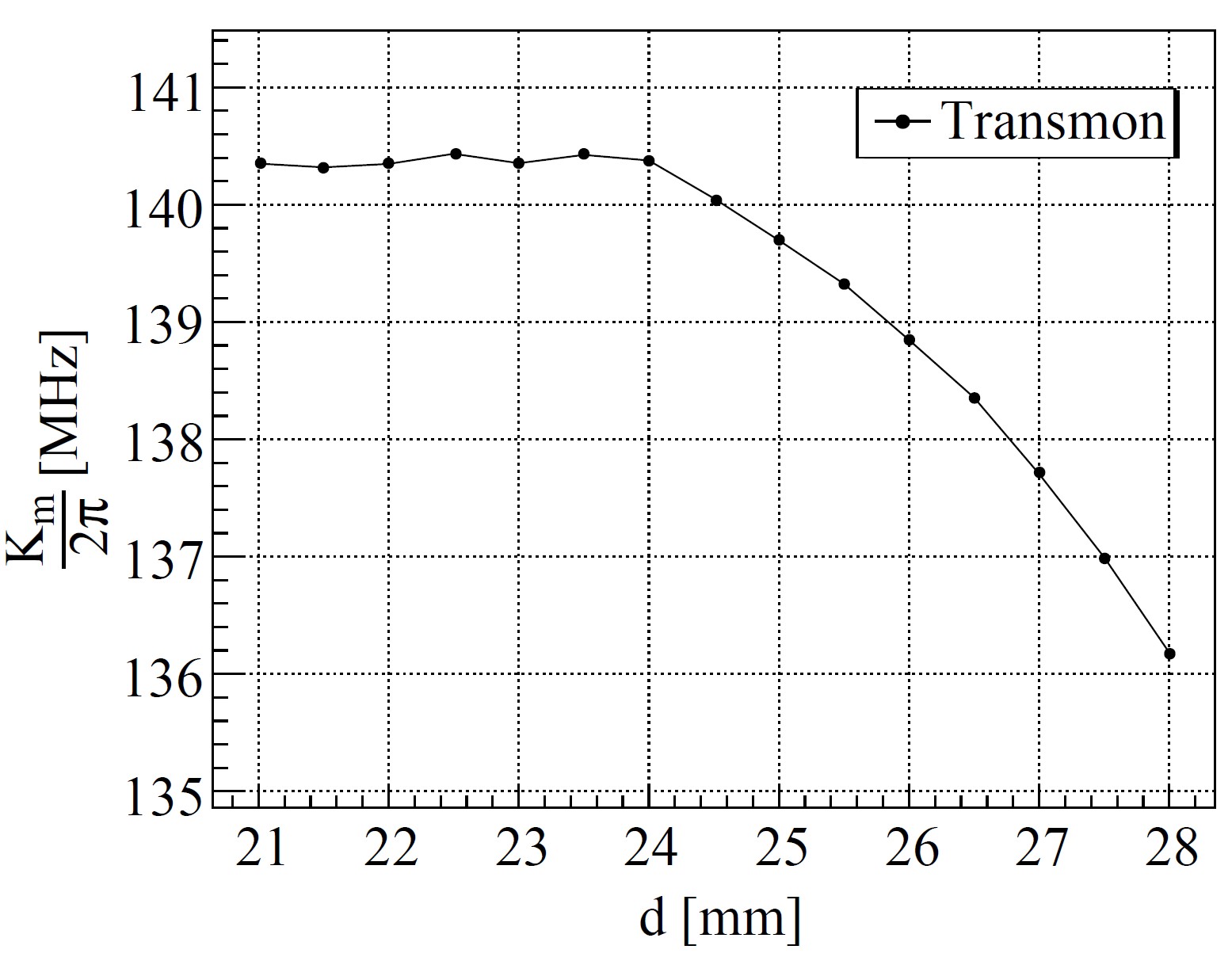}
       \label{fig:3CellAnharmVsPos}
    \end{subfigure}
    \caption{ Absolute values of the self-Kerr interactions as a function of transmon position for (a) cavity's $\mathrm{TM_{010}}$ modes and (b) transmon.}
    \label{fig:TESLALikeTMandTransmonSelfK}
\end{figure}

 The second set of parameters we evaluate as a function of the transmon position is the self-Kerr set, including the transmon anharmonicity. Since, as expected, the anharmonicity is considerably larger than the cavity modes' self-Kerrs, we report their absolute values in a logarithmic plot (Fig.\hspace{0.1cm}\ref{fig:TESLALikeTMandTransmonSelfK}). As highlighted before for the cross-Kerr interactions, the self-Kerr parameters of the first two $\mathrm{TM_{010}}$ modes are very similar for every transmon position, whereas the third mode's self-Kerr is roughly one order of magnitude less than the other two throughout the whole position sweep. All the self-Kerr magnitudes and the transmon anharmonicity, although reported in absolute value in the plots, are negative for any transmon position and decrease in absolute value as the transmon is moved outward. The cavity modes' self-Kerr magnitudes are considerably less than the corresponding modes' cross-Kerr values. Consequently, their effect on the coupled system behavior is very weak per electromagnetic excitation of a cavity mode, as stated by equation\hspace{0.1cm}$\left(\ref{eq:MultiModeDispHam}\right)$.

\par
\begin{figure}[htbp]
    \centering
    \includegraphics[scale = 0.18]{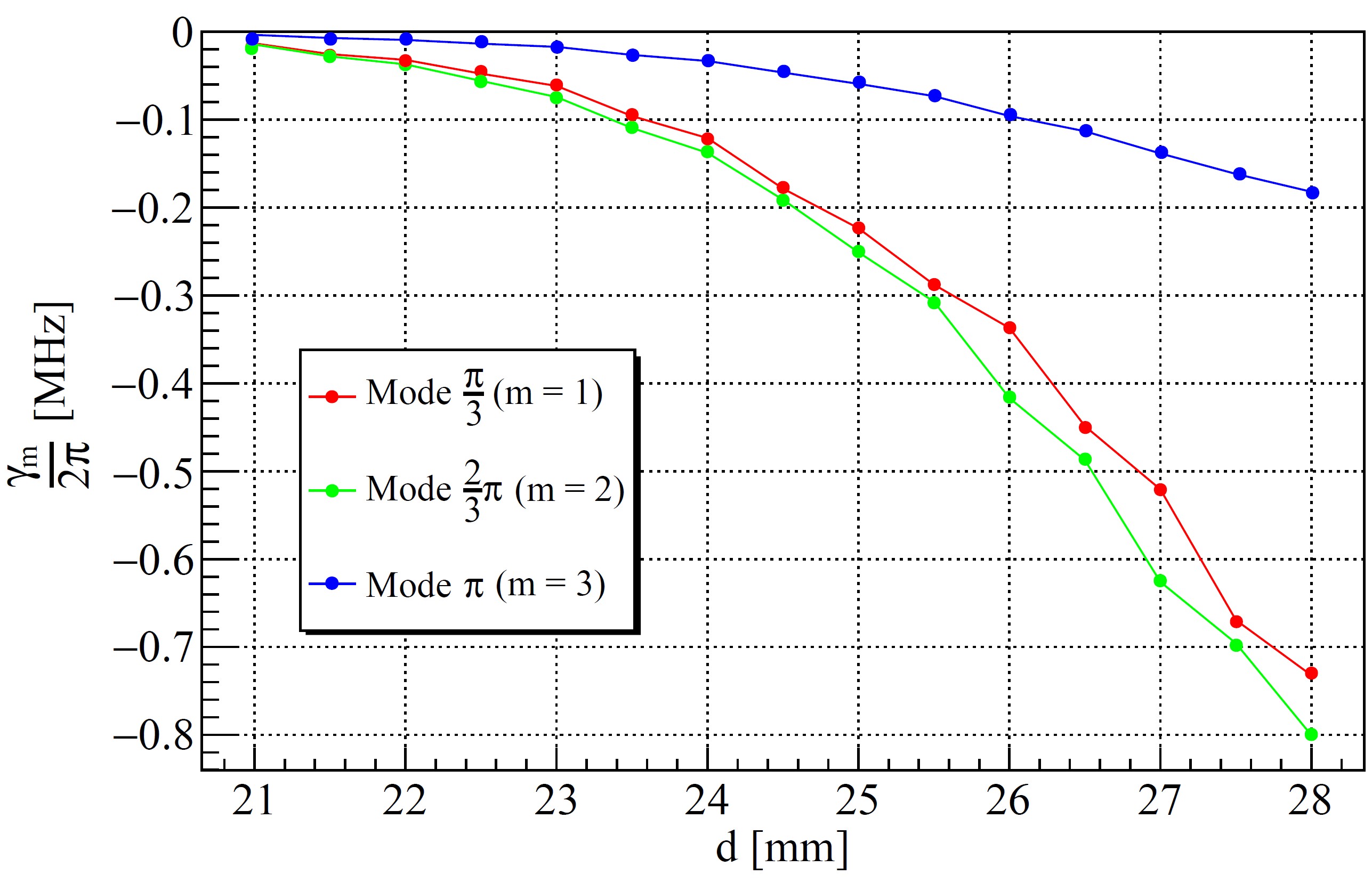}
    \caption{Linear correction magnitude to cavity modes as a function of transmon position for all cavity $\mathrm{TM_{010}}$ modes.}
    \label{fig:3CellLinCorrvsPosAllModes}
\end{figure}

 Proceeding with the energy-participation ratio analysis, the third ensemble of dispersive regime parameters we examine towards the transmon position is the linear corrections set. As for the self- and cross-Kerr cases, the first two $\mathrm{TM_{010}}$ modes' linear corrections are fairly similar throughout the entire position sweep (Fig.\hspace{0.1cm}\ref{fig:3CellLinCorrvsPosAllModes}). This supports the validity of Eq.~\eqref{eq:ParamRelationLinShifts} linking together all the dispersive regime parameters. Moreover, from the plot it is possible to notice that the corrections are almost equal to half the sum of self-Kerr and transmon-mode cross-Kerr magnitudes for each mode, implying that the cross-Kerr interactions between the cavity modes are negligible.

\par
\begin{figure}[htbp]
    \centering
    \includegraphics[scale = 0.18]{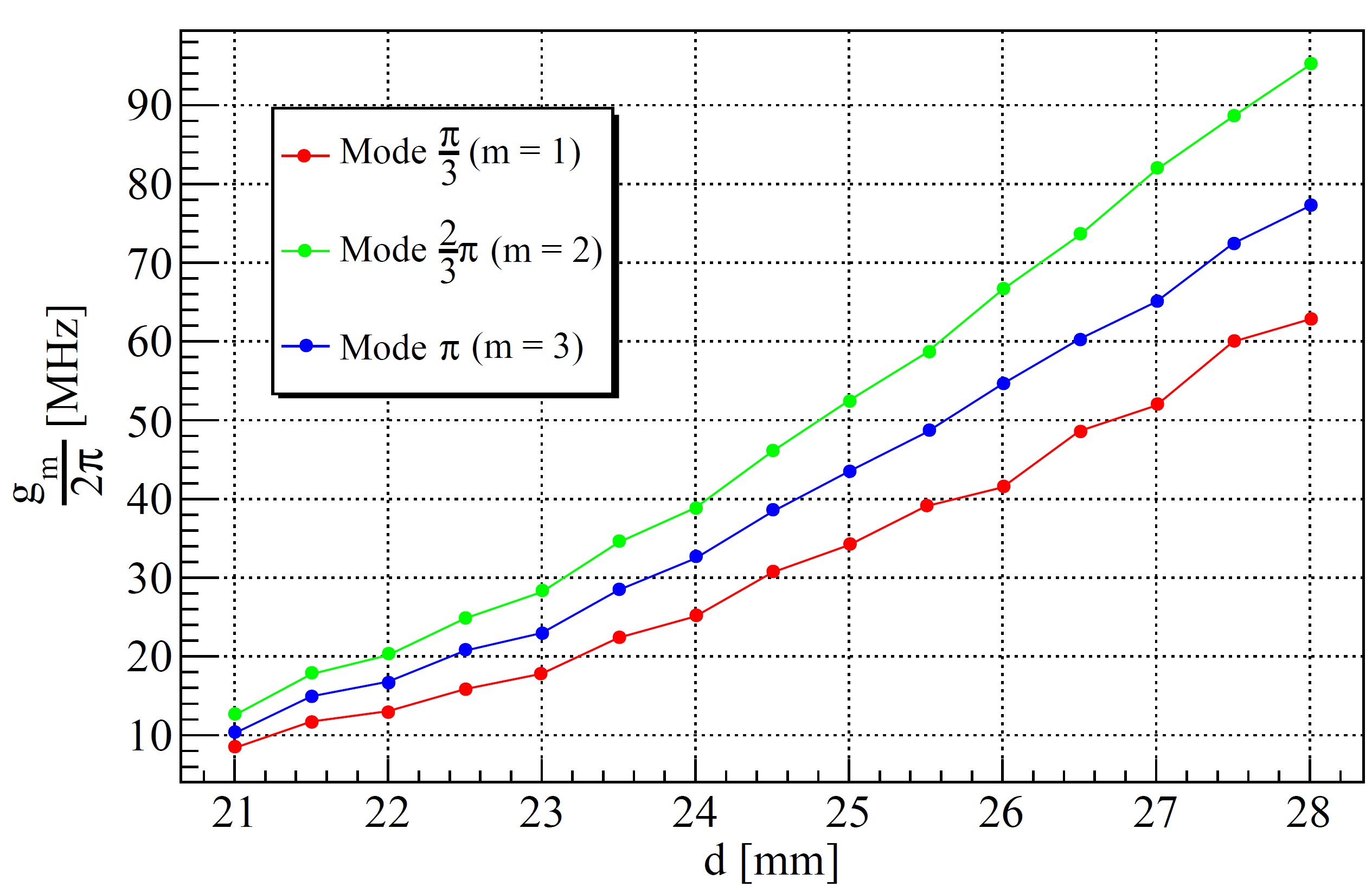}
    \caption{Rabi coupling magnitude as a function of transmon position for all $\mathrm{TM_{010}}$ modes.}
    \label{fig:3CellGvsPosAllModes}
\end{figure}

 Finally, the cross-Kerr interaction values are used to calculate the Rabi couplings through equation (\ref{eq:GVsdispPars}) and their behavior towards the transmon position is plotted. Contrarily to all the previous parameters, all the Rabi couplings are different from each other throughout the entire transmon position sweeps (Fig.\hspace{0.1cm}\ref{fig:3CellGvsPosAllModes}). However, this does not go against the similarities between the first two modes' parameters, as from equation (\ref{eq:GVsdispPars}) it can be seen that the difference in frequency $\Delta_{m}$ can compensate the $g_{m}$ variation, yielding an almost equal cross-Kerr magnitude for the two mentioned modes. This fact also justifies the difference in the parameters' magnitude characterizing the $\pi$-mode: its Rabi coupling is not enough to compensate for the frequency difference towards the transmon. Consequently, the $\pi$-mode parameters of cross-Kerr, self-Kerr and linear correction are smaller than the ones characterizing the other $\mathrm{TM_{010}}$ modes' interactions.

\section{Conclusions}
To summarize, we optimize the geometry of a high-coherence multi-mode cavity to improve its performance as a quantum processor. First of all, we systematically broaden the spectral width of the fundamental $\mathrm{TM_{010}}$ band, consequently increasing the spacing between modes and potentially resolving the issue of frequency crowding. Afterward, we establish a qualitative connection between some of the cavity geometric parameters and the electric field distribution of the fundamental modes. The results allow for more efficient optimization of the cavity design. The parameters found with the transmon position sweep in the second set of simulations allow a certain flexibility in the driving protocol choice, modifying the transmon-cavity interaction from strongly dispersive to weakly dispersive while retaining the similar cavity design. Future work will focus on the experiments to test the developed geometry with a prototype TESLA-like 3-cell cavity, which will provide guidance to further scale up the quantum processor.

\end{document}